\documentclass[twocolumn,showpacs,preprintnumbers,amsmath,amssymb]{revtex4}

\usepackage{graphicx}
\usepackage{dcolumn}
\usepackage{bm}
\usepackage{dsfont}
\usepackage{mathrsfs}
\usepackage{amsmath}				

\newcommand{\be}{\begin{equation}}
\newcommand{\ee}{\end{equation}}
\newcommand{\bey}{\begin{eqnarray}}
\newcommand{\eey}{\end{eqnarray}}

\begin{document} 
\title{Gauge and matter fields as surfaces and loops  - \\
an exploratory lattice study of the Z$_3$ Gauge-Higgs model} 
\author{Christof Gattringer, Alexander Schmidt}
\affiliation{
\vspace{3mm}
Institut f\"ur Physik, FB Theoretische Physik,
Universit\"at Graz, 8010 Graz, Austria
}

\date{September 1, 2012}

\begin{abstract}
\vspace{3mm}
We discuss a representation of the Z$_3$ Gauge-Higgs lattice field theory at finite
density  in terms of dual variables, i.e., loops of flux and surfaces. In the dual
representation the complex action problem of the conventional formulation is resolved and Monte Carlo simulations at
arbitrary chemical potential become possible. A suitable algorithm based on 
plaquette occupation numbers and
link-fluxes is introduced and we analyze the model at zero
temperature and finite density both in the weak and strong coupling phases.  We show
that at zero temperature the model has different 
first order phase transitions as a function of the
chemical potential both for the weak- and strong
coupling phases. The exploratory study demonstrates that alternative degrees of freedom may 
successfully be used for Monte Carlo simulations in several systems with gauge and matter fields. 
\end{abstract}

\preprint{INT-PUB-12-045}

\maketitle
 
\section{Introductory remarks}
In the last three decades lattice QCD has seen an impressive development into a
reliable tool for obtaining non-perturbative results in hadron physics. However,
for one important application, the study of QCD at finite density, the lattice
formulation has so far  not lived up to our expectations. The reason is that for
non-zero chemical potential the  fermion determinant is complex and cannot be
interpreted as a probability in a Monte Carlo simulation. Various approaches to
circumvent this problem were explored  and the reviews at the annual lattice
conferences provide a regular update \cite{reviews1,reviews2}.

An interesting approach to lattice systems with a complex action problem is to
search for a  mapping to alternative variables where the partition sum is a sum
over real and positive contributions such that in terms of the new variables the
complex action problem is  gone. Even for theories without a complex action problem
an alternative representation of the partition sum could allow for improved Monte
Carlo simulations with new algorithmic ideas. A prominent simple example is the
worm algorithm \cite{worm} for spin  systems such as the Ising model which is
based on a representation of the partition sum in terms of closed loops of 
Z$_2$ flux. 

In recent years several examples for new algorithmic approaches that are based on
transformations of lattice field theories to new variables were presented for a
wide range of applications: Low dimensional systems \cite{lowdim}, strongly
coupled lattice field theories \cite{strong}, systems with 4-fermi interactions
\cite{4fermi}, effective theories for QCD \cite{effqcd1,effqcd2}, scalar field
theories \cite{scalar1,scalar2} and U(1) lattice gauge theory
\cite{u1lgt1,u1lgt2}. Similar in spirit, simulations directly  based on the
Trotter formula for lattice field theories in Hamiltonian approach were  explored
\cite{trotter}. Many interesting conceptual and algorithmic ideas emerged in 
these papers and systems that previously were not fully accessible to Monte Carlo
simulations  can now be explored. 

So far the studies with alternative variables (dual variables) were mostly concerned  with flux-like
structures living on links, with the exception of the studies \cite{u1lgt1,u1lgt2}
of pure U(1) gauge theory where the dual variables are surfaces. In theories
where  gauge fields interact with matter fields, such as QCD, QED or Gauge-Higgs
systems a dual representation will contain both: Surfaces for the gauge fields and
fluxes for matter fields which serve as boundaries for open surfaces. 

In this article we develop the idea of using dual representations for simulations
of lattice field theories further and explore dual representations for the
Z$_3$ Gauge-Higgs systems. This is a first step towards systems which couple 
gauge- and matter fields and where surfaces
interacting with fluxes appear in the dual representation. As a matter of  fact 
all abelian Gauge-Higgs theories have a dual representation similar to the one we
here discuss for the Z$_3$ case and the techniques presented here can be
adapted to other abelian cases. The choice to use the gauge group Z$_3$ in this
exploratory study is partly motivated by the possibility to couple a chemical
potential and to explore finite density physics for Z$_3$, which as the center
group of SU(3) plays an interesting role in the phenomenology of QCD.     
At zero temperature we explore the various phase transitions as a function of 
the chemical potential. The main goal of this work is to develop further alternative 
representations of lattice field theories and their use in Monte Carlo simulations.  

In the next section we derive the dual representation of the partition sum in
terms of fluxes and surfaces. In Section III we discuss observables and
develop our strategy for the Monte Carlo simulation in terms of the dual
variables. In Section IV we first compare the dual simulation of pure Z$_3$ lattice
gauge theory and the full system at $\mu = 0$ to results from a Monte Carlo
calculation in the conventional approach. This is followed by  the presentation of
the results at finite density (Section V). A summary in Section VI completes the
paper.

\section{$\!\!$Dual representation of the Z$_3$ Gauge-Higgs model}
In this section we discuss the derivation of the dual representation for the 
Z$_3$ Gauge Higgs model on the lattice. Both the Higgs- and the gauge variables 
are elements of the group Z$_3 = \{1,e^{i2\pi/3}, e^{-i\pi/3} \}$, and it is
possible to couple a chemical potential $\mu$. In the conventional
representation the model then has a complex action problem at $\mu > 0$. 

The Higgs field variables $\phi_x$ live on the sites of a $N_s^3 \times N_t$
lattice  and are parameterized as  $\phi_x = e^{i s_x 2 \pi/3} $ with $s_x \in
\{-1,0,+1\}$. The gauge fields live on the links   and are written as
$U_{x,\sigma} = e^{i a_{x,\sigma} 2\pi/3}$ with $a_{x,\sigma}  \in \{-1,0,+1\}$.
For both fields periodic boundary conditions are used in all 4 directions. The
action $S = S_G + S_H$ is split into a gauge action $S_G$ and the action for the
Higgs field $S_H$. For the gauge action we use Wilson's form (without constant
term),
\begin{equation} 
S_G \; = \; - \frac{\beta}{2} \sum_x \sum_{\sigma < \tau} 
\Big[ U_{x,\sigma\tau} + U_{x,\sigma\tau}^\star \Big] \; ,
\label{gaugeactionz3}
\end{equation}
where the double sum runs over all plaquettes $U_{x,\sigma\tau} = U_{x,\sigma}
U_{x+\widehat{\sigma}, \tau} U_{x+\widehat{\tau},\sigma}^\star U_{x,\tau}^\star$.
The action for the Higgs field in a gauge field background configuration is  
\begin{equation}
S_H =  - \eta\!\! \sum_{x,\nu} \! \left[ e^{\mu \delta_{\nu,4}} \phi_x^\star \, U_{x,\nu} \,\phi_{x+\widehat{\nu}}   +
e^{- \mu \delta_{\nu,4}}  \phi_x^\star \, U_{x-\widehat{\nu}, \nu}^\star \, \phi_{x-\widehat{\nu}}  \right] \! \!  .
\label{higgsactionz3}
\end{equation}
The hopping terms in the 4-direction (= temporal direction) 
are coupled to the chemical potential $\mu$. The factor
$\eta$ controls the coupling between the Higgs- and the gauge fields.
Essentially it plays the role of an inverse Higgs mass: If the Higgs field is
infinitely heavy ($\eta = 0$) it  decouples from the dynamics of the system. The
partition sum is obtained by summing the Boltzmann factor over all possible
configurations
\begin{equation}
Z \; = \; \sum_{\{s,a\}} e^{-S_G - S_H}  \; ,
\end{equation}
with
\begin{equation}
\sum_{\{s,a\}} = \left(\prod_x \sum_{s_x = -1}^{1} \right)\!\left(\prod_{x,\sigma} 
\sum_{a_{x,\sigma} = -1}^1 \right).
\end{equation}
For deriving the dual representation of the Higgs field action we use two
identities which may be checked by explicit evaluation of both sides of the
equations for the
three cases $s = -1,0,1$. For the spatial nearest neighbor terms without chemical
potential we use ($ s = -1,0,1 $)
\begin{eqnarray}
&& \exp\left(\eta   e^{i\frac{2\pi}{3} s} + \eta e^{-i\frac{2\pi}{3} s} \right) \; = \; 
C_\eta \sum_{k=-1}^1 B_\eta^{|k|} \, e^{i \frac{2\pi}{3} s k} \;, 
\nonumber
\\
&& C_\eta \; = \; \frac{e^{2\eta} + 2 e^{-\eta}}{3} \quad , \qquad 
B_\eta \; = \; \frac{e^{2\eta} - e^{-\eta}}{e^{2\eta} + 2 e^{-\eta} } \; .
\label{aux1}
\end{eqnarray}
For the temporal hops where the chemical potential enters we need ($ s = -1,0,1 $)
\begin{eqnarray}
&& \exp\left(\eta e^\mu e^{i\frac{2\pi}{3} s} + \eta e^{-\mu} e^{-i\frac{2\pi}{3} s} \right) \; = \; \sum_{k=-1}^1 M_k \, e^{i \frac{2\pi}{3} s k}  \; ,
 \nonumber
\\
&&
M_k \; = \; \frac{1}{3} \Big[ e^{2\eta \cosh(\mu)} \label{aux2}  \\
&& \hspace{15mm} + \; 2 e^{-\eta \cosh(\mu) } \cos\Big( \sqrt{3} \eta \sinh(\mu) - k \frac{2\pi}{3}\Big) \Big] \; .
\nonumber
\end{eqnarray}
These auxiliary formulas are now used in an expansion of the Boltzmann factor. 
We find (the index $j$ runs from 1 to 3, $V = N_s^3$ is the spatial volume and we
use the representations $\phi_x = e^{i s_x 2 \pi/3} $ and $U_{x,\nu} = e^{i
a_{x,\nu} 2\pi/3}$)
\begin{eqnarray}
\hspace*{-35mm}&& \prod_{x,\nu} \exp\Big( \eta \, e^{\mu \, \delta_{\nu,4}} e^{i\frac{2\pi}{3} [ s_{x+\widehat{\nu}}  - s_x + a_{x,\nu} ] }  
\nonumber \\
&& \hspace{20mm} + \;
\eta \, e^{- \mu \, \delta_{\nu,4}} e^{- i\frac{2\pi}{3} [  s_{x+\widehat{\nu}}  - s_x + a_{x,\nu} ] }  \Big)
\nonumber \\
\hspace*{-35mm} && =  \; C_\eta^{3V} \sum_{\{k\}}  \!\! \left( \! \prod_{x,j}\! B_\eta^{|k_{x,j}|} \, e^{i\frac{2\pi}{3} 
\big[ s_{x+\widehat{j}}  - s_x + a_{x,j} \big] k_{x,j} }\! \right)
\nonumber \\
&& \hspace{18mm} \times \; 
\left(\! \prod_x\! M_{k_{x,4}} \, e^{i\frac{2\pi}{3} \big[ s_{x+\widehat{4}}  - s_x + a_{x,4} \big] k_{x,4} } \!\right) .
\qquad \;
\end{eqnarray}
We have introduced the shorthand notation $\sum_{\{k\}} = \prod_{x,\nu}
\sum_{k_{x,\nu}=-1}^1$ for the sum over all configurations of the expansion
indices $k$.  Inserting the expanded Boltzmann factor into the partition sum for
the Higgs field in a given gauge field background (represented by the coefficients
$a_{\nu,x}$) we find
\begin{eqnarray}
&& \hspace{-3mm} Z_H[a]  =  \sum_{\{s\}} e^{-S_H} = C_\eta^{3V} \sum_{\{k\}} \left( \prod_{x,j} B_\eta^{|k_{x,j}|} \right)\!\!\left( \prod_x M_{k_{x,4}} \right)
\nonumber 
\\
&& \hspace{-3mm}  \times \! \left (\prod_{x,\nu} e^{i \frac{2\pi}{3} a_{x,\nu} k_{x,\nu} } \right) \!\!
 \left( \prod_x \! \sum_{s_x = -1}^1 e^{- i \frac{2\pi}{3} s_x \sum_\nu \big[ k_{x,\nu} - k_{x-\widehat{\nu},\nu} \big] } \! \right) \! ,
\nonumber \\ 
\end{eqnarray}
where we have suitably reorganized the product over the link terms. The product of
sums in the last term of this expression is the remaining sum over the
configurations of the Higgs fields (parameterized by $s_x$). Each of the
individual sums over the $s_x$ vanishes, unless  $\sum_\nu \big[ k_{x,\nu} -
k_{x-\widehat{\nu},\nu} \big]$ is a multiple of 3. It is useful to define the
triality function $T(n)$ as
\begin{equation}
T(n) \; = \; \left\{ 
\begin{array}{cc}
1 & \mbox{if} \;\; n \; \mbox{mod} \, 3 \; = \; 0 \; ,
\\
0 & \mbox{if} \;\; n \; \mbox{mod} \, 3 \; \neq \; 0 \; .
\end{array} \right.
\label{triality}
\end{equation}
Using the triality function $T(n)$ we write the Higgs-field partition sum as
\begin{eqnarray}
\hspace*{-4mm}
Z_H[a]  &=& C_\eta^{3V}3^V \sum_{\{k\}} \left( \prod_{x,j} B_\eta^{|k_{x,j}|} \right)\!\!\left( \prod_x M_{k_{x,4}} \right)
\nonumber\\
&&\hspace{-15mm}  \times \! \left (\prod_{x,\nu} e^{i \frac{2\pi}{3} a_{x,\nu} k_{x,\nu} } \! \right) \!\!
\left(\!\prod_x T\Big(\sum_\nu\big[k_{x,\nu} - k_{x-\widehat{\nu},\nu}\big]\Big)\!\!\right)\! .
\label{zhz3}
\end{eqnarray}
The last factor implements a constraint at each site 
$x$, i.e., the net-flux through $x$, given by 
$\sum_\nu\big[k_{x,\nu} - k_{x-\widehat{\nu},\nu}\big]$,  
has to vanish modulo 3.

If the net flux through a site were to vanish exactly (not only modulo 3) the
constraint would imply that only closed loops of flux are allowed. The fact that the
net flux through each site vanishes only modulo 3 augments the closed loop condition
with an additional rule allowing 3 units of flux to emerge from a site or to vanish
at a site. 

The next step is to use the flux representation (\ref{zhz3}) for the Higgs part in
the expression for the full partition sum. The Boltzmann factor for the gauge part
is also expanded, re-using the identity (\ref{aux1}) in the form ($a = -1,0,1$)
\begin{eqnarray}
&& \exp\left(\frac{\beta}{2}   e^{i\frac{2\pi}{3} a} + \frac{\beta}{2} e^{-i\frac{2\pi}{3} a} \right) \; = \; 
C_\beta \sum_{p=-1}^1 B_\beta^{|p|} \, e^{i \frac{2\pi}{3} a p}  \; ,
\nonumber \\
&& C_\beta \; = \; \frac{e^{\beta} + 2 e^{-\frac{\beta}{2}}}{3} \quad , \qquad 
B_\beta \; = \; \frac{e^{\beta} - e^{-\frac{\beta}{2}}}{e^{\beta} + 2 e^{-\frac{\beta}{2}} } \; .
\label{aux3}
\end{eqnarray}

Representing the Boltzmann factor for the gauge fields with this formula and
inserting it in the full partition sum we find after some reordering of terms
\begin{eqnarray}
\hspace*{-4mm}&& Z  =  \sum_{\{a\}} e^{-S_G} Z_H[a]  =   C_\eta^{3V}3^V C_\beta^{6V} \!\!\sum_{\{p,k\}}  \!\!\!
\left( \prod_{x,j} B_\eta^{|k_{x,j}|}\! \right)\!\! \times
 \label{aux7}   \\
\hspace*{-4mm}&& 
\left(\! \prod_x M_{k_{x,4}} \!\right)\!\! \left(  \prod_{x,\sigma < \tau} B_\beta^{|p_{x,\sigma\tau}|}\!\right) \!\!
\left(\!\prod_x T\Big(\sum_\nu\big[k_{x,\nu} - k_{x-\widehat{\nu},\nu}\big]\Big)\!\!\right)
\nonumber \\  
\hspace*{-4mm}&&\qquad \times
\Bigg(\!\prod_{x,\nu}  \sum_{a_{x,\nu}=-1}^1 \!\!\!\!\! \exp\! \Bigg(\! i\frac{2\pi}{3}  a_{x,\nu} \times
\nonumber \\
\hspace*{-4mm}&&\!\!
\Bigg[\! \sum_{\nu < \alpha}\! \big[ p_{x,\nu\alpha} - p_{x-\widehat{\alpha},\nu\alpha}\big]
- \sum_{\alpha<\nu} \! \big[ p_{x,\alpha\nu} - p_{x-\widehat{\alpha},\alpha\nu}\big]
  + k_{x,\nu}\! \Bigg] \!  \Bigg) \!\! \Bigg) .
\nonumber
\end{eqnarray}
We have introduced plaquette occupation numbers $p_{x,\sigma\tau}$ which may
assume the values $-1,0$ and $+1$ and for unique labeling of the plaquettes we only consider
$p_{x,\sigma\tau}$ with $\sigma < \tau$. As before, by $\sum_{\{p\}}$ we denote the sum
over all configurations of the $p_{x,\sigma\tau}$. The last product of sums  in
(\ref{aux7}) gives rise to triality constraints for all links which combine the
plaquette occupation numbers of all plaquettes attached to that link and the
$k$-flux residing on that link. In a more compact notation the final result for
the partition sum of the Z$_3$ Gauge-Higgs model reads
\begin{equation}
Z \; = \; C \sum_{\{p\}}  \sum_{\{k\}} {\cal C}_P[p,k] \,  {\cal C}_F[k] \,  {\cal W}_P[p] \,  {\cal W}_F[k] \; .
\label{ZfinalZ3}
\end{equation} 
The first sum runs over all configurations of the integer valued plaquette
occupation variables $p_{x,\sigma\tau} \in \{-1,0,+1\}$ assigned to the plaquettes
of the lattice, while the second sum is over all configurations of the flux
variables  $k_{x,\nu} \in \{-1,0,+1\}$ living on the links of the lattice. The
flux variables $k$ are subject to the constraint $ {\cal C}_F[k]$ given by
\begin{equation}
 {\cal C}_F[k] \; = \; \prod_x T \left( \sum_\nu \big[ k_{x,\nu} - k_{x-\widehat{\nu},\nu} \big] \right) \; ,
\end{equation}
which enforces the conservation of $k$-flux modulo 3 at each site of the lattice
(see (\ref{triality}) for the definition of $T(n)$). This flux conservation
restricts the admissible configurations to closed oriented loops of $k$-flux. A
second constraint,
\begin{eqnarray}
 && {\cal C}_P[p,k]  =  \prod_{x,\nu} T  \bigg( \sum_{\nu < \alpha}\! \big[ p_{x,\nu\alpha} - p_{x-\widehat{\alpha},\nu\alpha}\big] \qquad
\label{plaqconstZ3}  \\
&& \hspace{25mm}  - \sum_{\alpha<\nu}\! \big[ p_{x,\alpha\nu} - p_{x-\widehat{\alpha},\alpha\nu}\big]  + k_{x,\nu} \!\bigg)\!  ,
\nonumber
\end{eqnarray}
connects the plaquette occupation numbers $p$ with the $k$-variables: At every
link it enforces the combined flux of the plaquette occupation  numbers attached
to that link plus the $k$-flux on that link to vanish modulo 3. Similar to the
closed loop interpretation for the $k$-flux, the link constraint   forces the
plaquette occupation numbers $p$ into forming closed surfaces, or open surfaces
bounded by loops of $k$-flux, and again the surface rule is augmented by the 
modulo 3 exemption as for the fluxes. 

Both, the plaquette occupation numbers and the fluxes come with corresponding
weight factors, given by
\begin{eqnarray}
 {\cal W}_P [p] &\!\!\! = \!\!\! & \prod_{x,\sigma < \tau} B_\beta^{|p_{x,\sigma\tau}|} \; ,
\label{Z3gaugeweight} 
\\
 {\cal W}_F [k] &\!\!\!  = \!\!\! &\left( \prod_{x,j}\!  B_\eta^{|k_{x,j}|}\right)\!\left( \prod_x M_{k_{x,4}}\right)  ,
\nonumber
\end{eqnarray}
where the explicit expressions for $B_\eta$ and $B_\beta$ are given in
(\ref{aux1}) and (\ref{aux3}). The partition sum  (\ref{ZfinalZ3}) comes with an
overall normalization factor $C = (3^5 C_\eta^3 C_\beta^6)^V$ where $V = N_s^3$
and $C_\eta$ and $C_\beta$ are given in (\ref{aux1}) and (\ref{aux3}). 

\section{Observables and Monte Carlo update}
 
Having mapped the partition sums onto the dual variables we now also need to
identify the representation of the observables in terms of the dual degrees of
freedom. In this exploratory study we concentrate on thermodynamical observables,
i.e., observables that are obtained as derivatives of $\ln Z$ with respect to the
various couplings \footnote{It is straightforward to derive the dual representation
for several correlators and also for Wilson loops by introducing locally varying
couplings, taking suitable derivatives and then setting the couplings back to the
values one wants to work at. In our exploratory analysis of the
Z$_3$ Gauge-Higgs systems with dual variables
we do not present results
for this type of observables.}. For the gauge sector we obtain the average plaquette
$\langle U \rangle$ and the corresponding susceptibility  $\chi_U$ as first and
second derivatives with respect to the inverse gauge coupling $\beta$,
\begin{equation}
\langle U \rangle =  \frac{1}{6 N_s^3 N_t} \, \frac{\partial}{\partial \beta} \, \ln Z \; ,
\; \chi_U  = \frac{1}{6 N_s^3 N_t} \, \frac{\partial^2}{\partial \beta^2} \, \ln Z \; .
\label{obs1}
\end{equation}
Both observables are normalized by the total number of plaquettes. The
$\beta$-dependence of the dual representation is encoded in the weight factors 
given by the product (\ref{Z3gaugeweight})  over powers $B_\beta^{|p|}$ of the factors $B_\beta$
defined in (\ref{aux3}).  There is an additional $\beta$-dependence from the
overall factor $C$ which also contains $B_\beta$. This factor is necessary for the correct representation 
of the observables listed in (\ref{obs1}).  

It is straightforward to evaluate the expressions (\ref{obs1}) in the dual 
representation and one obtains:
\begin{eqnarray}
\hspace*{-4mm} && \langle U \rangle  =  B_\beta  \; + \;  \frac{1}{6 N_s^3 N_t} \frac{B_\beta^\prime}{B_\beta}   
\left \langle \sum_{x,\sigma < \tau} | p_{x,\sigma \tau} | \right \rangle \; ,
\label{obs1Z3}
\\
\hspace*{-4mm} && \chi_U  =   B_\beta^\prime  \; + \;     \frac{1}{6 N_s^3 N_t} \, \frac{B_\beta^{\prime\prime} \, B_\beta - 
\big(B_\beta^\prime\big)^2}{B_\beta^2}
\left \langle \sum_{x,\sigma < \tau} | p_{x,\sigma \tau} | \right \rangle  +
\nonumber \\
\hspace*{-4mm} && \frac{1}{6 N_s^3 N_t} \left( \frac{B_\beta^\prime}{B_\beta}\right)^{\!2} \!   \left[   \left \langle \! \left( \sum_{x,\sigma < \tau} | p_{x,\sigma \tau} |  \right)^{\!\!2} 
\right \rangle 
- \left \langle \sum_{x,\sigma < \tau} | p_{x,\sigma \tau} | \!  \right \rangle^{\!\!2} \right] \! .
\nonumber
\end{eqnarray}
 
For the Higgs sector we study the particle number density $n$ and the corresponding susceptibility 
$\chi_n$ which may
be obtained as derivatives with respect to $\mu$ (again normalized with the 4-volume):
\begin{equation}
n \; = \; \frac{1}{N_s^3 N_t}  \frac{\partial}{\partial \mu} \, \ln Z \; , \;
\chi_{n} \; = \; \frac{1}{N_s^3 N_t} \, \frac{\partial^2}{\partial \mu^2} \, \ln Z \; .
\end{equation}
The corresponding observables in the dual representation are
\begin{eqnarray}
n &\! =\! & \frac{1}{N_s^3 N_t} \big \langle R_1\, {\cal N}_1 \, + \, R_0\,  {\cal N}_0 \, + \,R_{-1} \,  {\cal N}_{-1} \big \rangle \; ,
\label{obs2Z3}
\\
\chi_n &\! =\! &  \frac{1}{N_s^3 N_t} \Big[ \big \langle Q_1\,  {\cal N}_1 \, + \, Q_0\,  {\cal N}_0 \, + \,Q_{-1} \,  {\cal N}_{-1} \big \rangle 
\nonumber \\
&\!+\! &\!  \big \langle \big( R_1\,  {\cal N}_1 \, + \, R_0\,  {\cal N}_0 \, + \,R_{-1} \,  {\cal N}_{-1} \big)^2 \big \rangle 
\nonumber \\
&\!-\! & \big \langle  R_1\,  {\cal N}_1 \, + \, R_0\,  {\cal N}_0 \, + \,R_{-1} \,  {\cal N}_{-1}   \big \rangle^2 \Big] .
\nonumber
\end{eqnarray}
We introduced the abbreviations ${\cal N}_s, \, s = -1,0,1$ for the total number of temporal link variables $k_{x,4}$ that have a value $s = -1,0,1$.
The $R_k$ and $Q_k$ with $k = -1,0,1$ denote the ratios $R_k = M_k^\prime/M_k$ and $Q_k = [M_k^{\prime \prime}M_k - (M_k^\prime)^2]/M_k^2$,
where the $M_k$ are the factors defined in (\ref{aux2}) and the primes are used for their derivatives with respect to $\mu$.

\vskip5mm

Although the partition sums and the observables may seem somewhat involved in the
dual representation, the dual Monte Carlo update turns out to be rather  simple. We
begin its discussion with introducing an update for pure gauge theory which in the
next section we study  as a first test case. The pure gauge update then serves as the
starting point for developing the full algorithm where we will augment the pure gauge
algorithm with another simple step to update the Higgs field as well.

In the dual representation the pure gauge theory is a theory of only the plaquette
variables and the partition sum can be written as (see (\ref{ZfinalZ3})), 
\begin{equation}
Z_G \; = \; \sum_{\{p\}} C_P[p] \, W_P[p] \; ,
\end{equation}
where the sum is over configurations of the plaquette occupation numbers
$p_{x,\sigma \tau} \in \{-1,0,1\}$. The constraint $C_P[p]$ is
simply the plaquette constraint (\ref{plaqconstZ3}) evaluated for the case
when all $k$-fluxes are set to $k_{x,\nu} =
0$. The constraint is a product over all links and for each link the oriented
flux of all plaquettes attached to that link has to be a multiple of 3 (see
(\ref{plaqconstZ3})). As discussed in the previous section the constraints
give rise to the occupied plaquettes forming surfaces. Here, where we consider pure
gauge theory, we have only closed surfaces since without matter fields there are no
$k$-fluxes that could serve as boundaries of open surfaces.

For the update of the plaquette variables we start from a configuration of the
plaquette occupation numbers $p_{x,\sigma \tau}$  where all the link constraints
are satisfied, using the simplest choice, i.e., the trivial configuration with
$p_{x,\sigma \tau} = 0$ for all $x$ and all $\sigma$, $\tau$. Starting from the
trivial configuration we offer trial configurations which leave the
constraints intact and accept them with the usual  Metropolis probability.  

\begin{figure}[t]
\centering
\includegraphics[width=0.45\textwidth,clip]{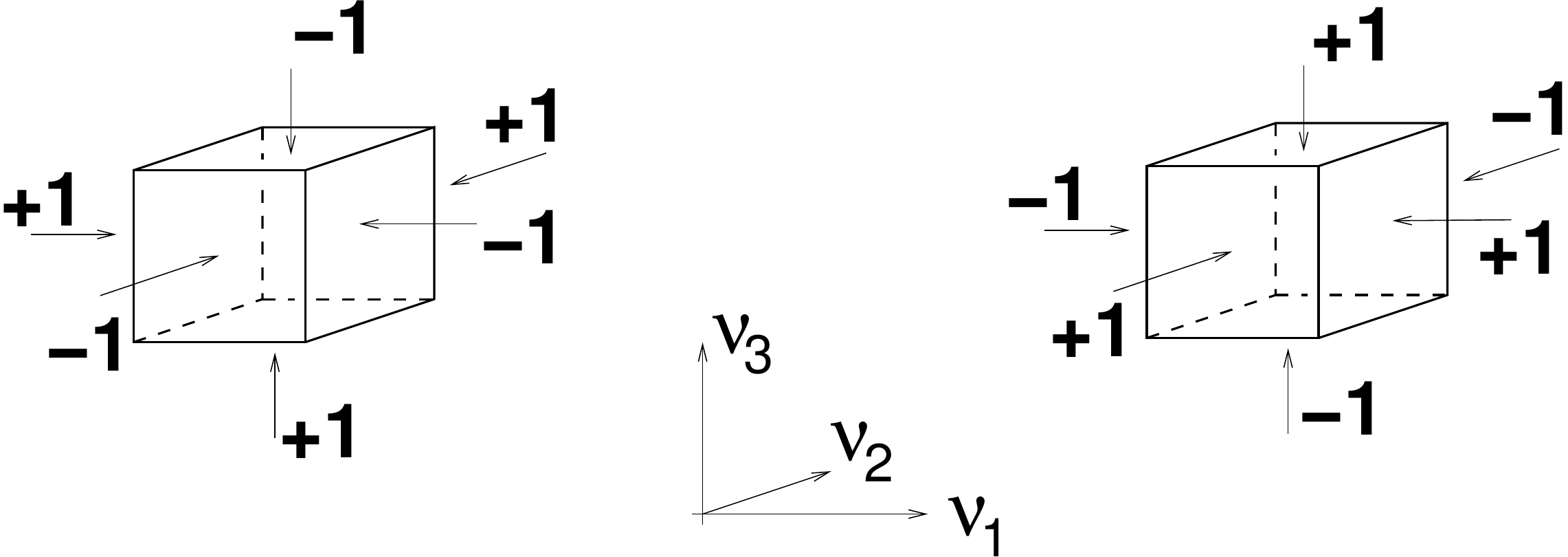}
\caption{Cube update: We illustrate the changes we propose for the plaquette occupation numbers at the faces of an embedded 3-cube
with edges along the directions  $1 \leq \nu_1 < \nu_2 < \nu_3 \leq 4$. The two choices in the lhs.\ and rhs.\ figure are proposed with equal probability.}
\label{cubeupdate}
\end{figure}

The trial configurations are generated by increasing or decreasing the plaquette
occupation numbers on the faces of 3-cubes which we embed in four dimensions. Let
$1 \leq \nu_1 < \nu_2 < \nu_3 \leq 4$ denote the three directions that define the
3-cube. Then there are four possible choices of the $\nu_i$, i.e., four different
3-cubes can be embedded in four dimensions. Once the three directions $\nu_i$ are
fixed we select a site $x$ for  the lower left front corner of the cube and change
the plaquette occupation numbers on the faces of the cube by $\pm 1$ according to
one of the two possibilities depicted in Fig.~\ref{cubeupdate}.

To take into account the fact that the plaquette occupation numbers are restricted
to $-1,\, 0$ and $+1$, the addition of $\pm 1$ is  understood only modulo 3.
Addition modulo  3 is defined as the usual addition of the numbers $+1$, $0$ and
$-1$, except for the two cases $+ 1 + 1 \equiv -1$ and $- 1 - 1 \equiv +1$.  It is easy to
check that the two changes illustrated in Fig.~\ref{cubeupdate} leave the
constraints at the sides of the cubes  intact. The two possible changes are
proposed with equal probability and are then accepted with the usual Metropolis
probability. The corresponding acceptance rate is a product of factors
$(B_\beta)^{ \pm 1}$ according to the plaquette occupation numbers at the faces of
the cubes (see Eq.~\ref{Z3gaugeweight}). A full cube  sweep is a loop over all four
possible embeddings of 3-cubes  and over the $N_s^3 \times N_t$ possibilities to
place a particular cube.

On a finite lattice with periodic boundary conditions there are additional
admissible configurations of the plaquette occupation numbers which are not
properly treated by the cube updates alone. They consist of surfaces that close
around the periodic boundaries. To take them into account we offer an additional
set of trial configurations, where we change all plaquette occupation numbers in a
plane by $\pm 1$ (addition is again understood modulo 3) and accept this change
with the corresponding Metropolis probability which  is a product of powers of
$B_\beta$ according to the plaquette occupation numbers on the plane of the trial
configuration. For a complete ''plane sweep'' we offer this change for all
possible planes (6 possible orientations of the planes with $N_s^2$ or $N_s N_t$
slices for each orientation). For the update of pure gauge theory we
mix plane and cube sweeps. 

\begin{figure}[t!]
\centering
\includegraphics[width=0.45\textwidth,clip]{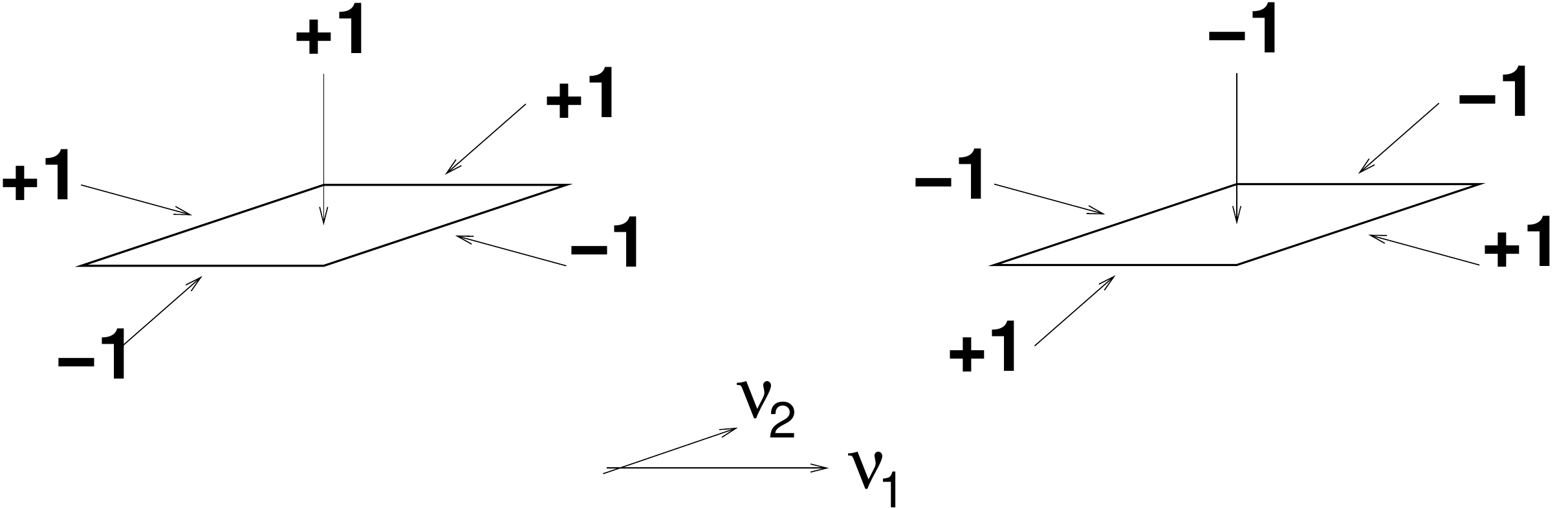}
\caption{Plaquette update: We illustrate the changes we propose for the plaquette occupation numbers at the plaquette and the corresponding 
changes of the flux variables at the links of the plaquette. The directions $1 \leq \nu_1 < \nu_2 \leq 4$ determine the plane of the plaquette and the 
two choices in the lhs.\ and rhs.\ figure are proposed with equal probability.}
\label{plaqupdate}
\end{figure}

For updating the full Gauge-Higgs systems also the matter fields need to be taken
into account. They are represented by the link variables. The update of the
$k$-fluxes is more interesting since they enter in two constraints: At each site
the total flux of the $k$-variables has to vanish modulo 3. Furthermore, at
each link the corresponding $k$-flux enters the constraint for the plaquette
occupation numbers attached to that link. To deal with this situation we use
proposal  configurations that change a plaquette occupation number and the
$k$-fluxes at the links of that plaquette. The two schoices for such a
change are  depicted in Fig.~\ref{plaqupdate}, where $1 \leq \nu_1 < \nu_2 \leq 4$
are the two directions that define the plane of the plaquette. 

It is easy to see that the two moves we propose do not change the  constraints on the
sites or on the links (addition is again only understood modulo 3 as for the cube and
plane updates). The two changes in Fig.~\ref{plaqupdate} are proposed with equal
probability and are accepted with the Metropolis acceptance rate. This is a  product of a
ratio of the plaquette weights  $B_\beta$ and a product over ratios of the 4 link weights,
i.e., powers of $B_\eta$ for the spatial links, and factors $M_k$ for the temporal links.
A full plaquette sweep then consists of a loop over all 6 embeddings of the planes of the
plaquettes and all $N_s^3 \times N_t$ possibilities to place the plaquette. 

To summarize: For updating the full Gauge-Higgs model we use combined sweeps
that mix full cube sweeps, full plane sweeps and 
full plaquette sweeps for the $k$-variables. 

It is interesting to note that when the Higgs field is coupled, i.e., for $\eta
> 0$, the cube and plane sweeps could be omitted and it would be sufficient to
work with  the plaquette sweeps alone. The reason is that a cube update can be
achieved by 6 plaquette updates, and a plane sweep by a combination of plaquette
updates in the respective plane. However, we found that in particular omitting the
cube updates may lead to a poorer performance of the algorithm in cases where the
weights for the $k$-fluxes are small. We remark, that for the Higgs model without
gauge fields a very efficient worm algorithm  \cite{worm} can be constructed, and
also for the full Z$_3$-case a surface type of  generalization of the worm idea is
possible \cite{ydalia} (see also \cite{u1lgt2} for tests in this direction).

\section{Comparison with conventional $\mu = 0$ results}

\subsection{Pure gauge theory}

Having discussed the dual representation and the Monte Carlo algorithm in the dual
picture  we now come to the evaluation of the dual Monte Carlo approach. As a warm
up exercise we  begin with the case of pure gauge theory, where the Monte Carlo
update is based on the cube  and plane updates. The observables we consider are
the plaquette  expectation value $\langle U \rangle$ and the corresponding
susceptibility $\chi_U$.  Their definitions and the corresponding 
dual expressions are given in 
(\ref{obs1}) and (\ref{obs1Z3}). 

\begin{figure*}[t]
\centering
\includegraphics[width=0.78\textwidth,clip]{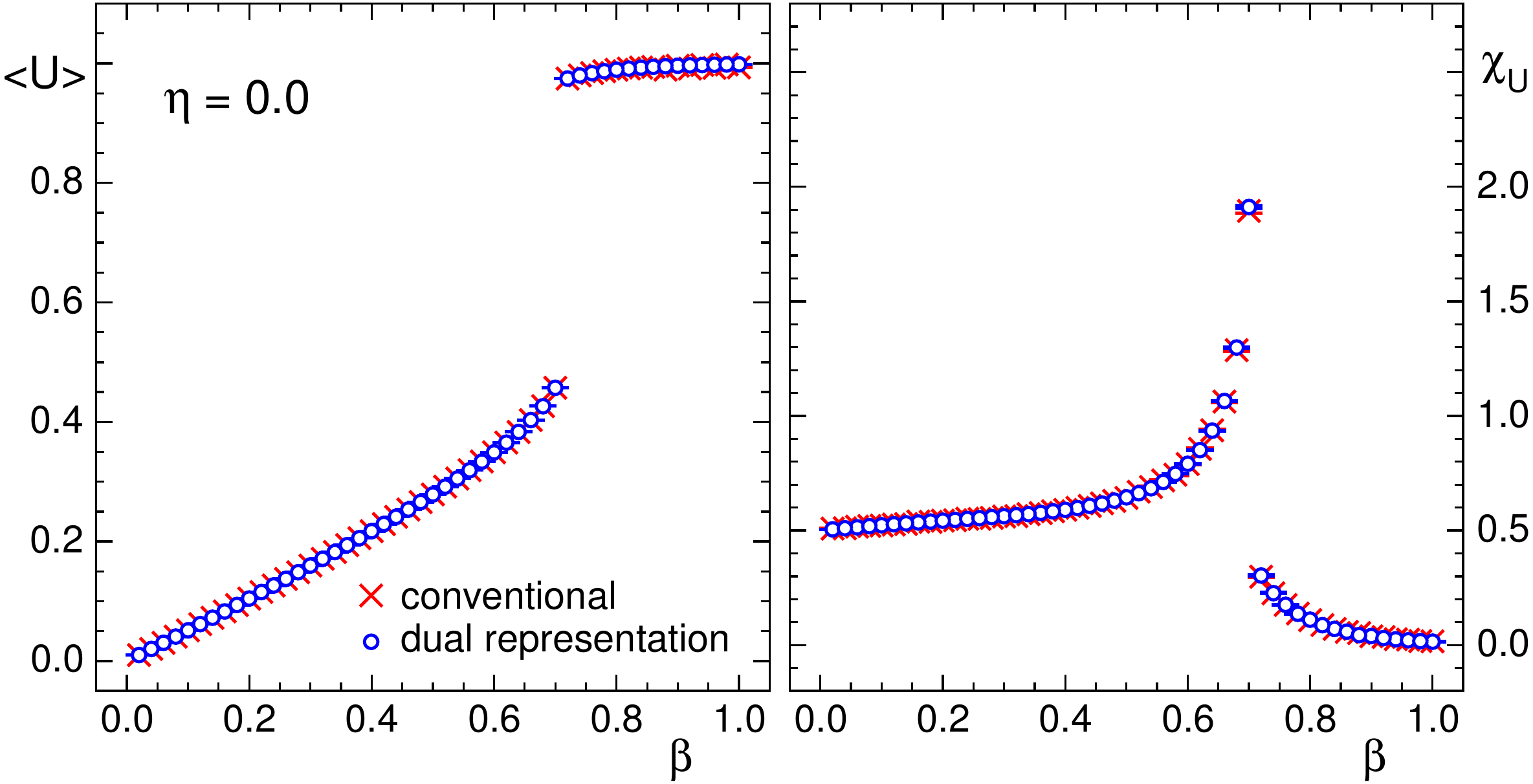}
\caption{Results for plaquette $\langle U \rangle$ (lhs.~plot) and the plaquette
susceptibility $\chi_U$ in pure Z$_3$ gauge theory as a function of the inverse
gauge coupling $\beta$. We compare the conventional (crosses) and the dual
approach (circles).}
\label{pure_gauge_Z3}
\end{figure*}

In Fig.~\ref{pure_gauge_Z3} we show the results for $\langle U \rangle$ and
$\chi_U$  in pure Z$_3$ lattice gauge theory ($\eta = 0$), and compare the
outcome of the dual  simulation (circles) to the results from the conventional
approach (crosses) using lattices with volumes of size $10^4$.  The statistics is 10000 configurations
separated by 10 cube sweeps and one plane sweep and another 10000 cube sweeps
mixed with 1000 plane sweeps were used for equilibration. All errors we show in
this  work are statistical errors determined with the jackknife method.

Fig.~\ref{pure_gauge_Z3} shows that the results of the dual simulation perfectly agree with the outcome of the
conventional approach. Near $\beta_c \sim 0.7$ the system apparently  undergoes a quite
prominent first order transition which separates the strong- ($\beta < \beta_c$)
and weak-coupling ($\beta > \beta_c$) phases, and also near the transition the
results  for the first and second derivatives of the free energy (i.e., $\langle U
\rangle$ and $\chi_U$) obtained with the conventional and dual  approaches agree
perfectly. We conclude that for the case of pure  Z$_3$ lattice gauge theory the
mapping to the dual representation, the identification of the observables and the
simulation  with the cube and plane algorithms work.

It is interesting to inspect the acceptance rate for the plane sweeps.  We find
that the acceptance is zero for the strong coupling phase, i.e., for $\beta <
\beta_c$. For a small volume of size $4^4$ we then see the onset of nonzero
acceptance at  $\beta_c  \sim 0.7$ reaching a value of  0.28 at $\beta = 1.0$.
Repeating the same analysis on volumes of size 
$10^4$ we again find non-zero acceptance
only above $\beta_c \sim 0.7$, but the increase with $\beta$ is much slower, and at 
$\beta = 1.0$ we still see an acceptance rate smaller than 0.01. We conclude that the
non-trivially winding sheets of occupation are a finite size effect that very quickly
dies out with increasing volume.

\begin{figure*}[t!]
\centering
\includegraphics[width=0.75\textwidth,clip]{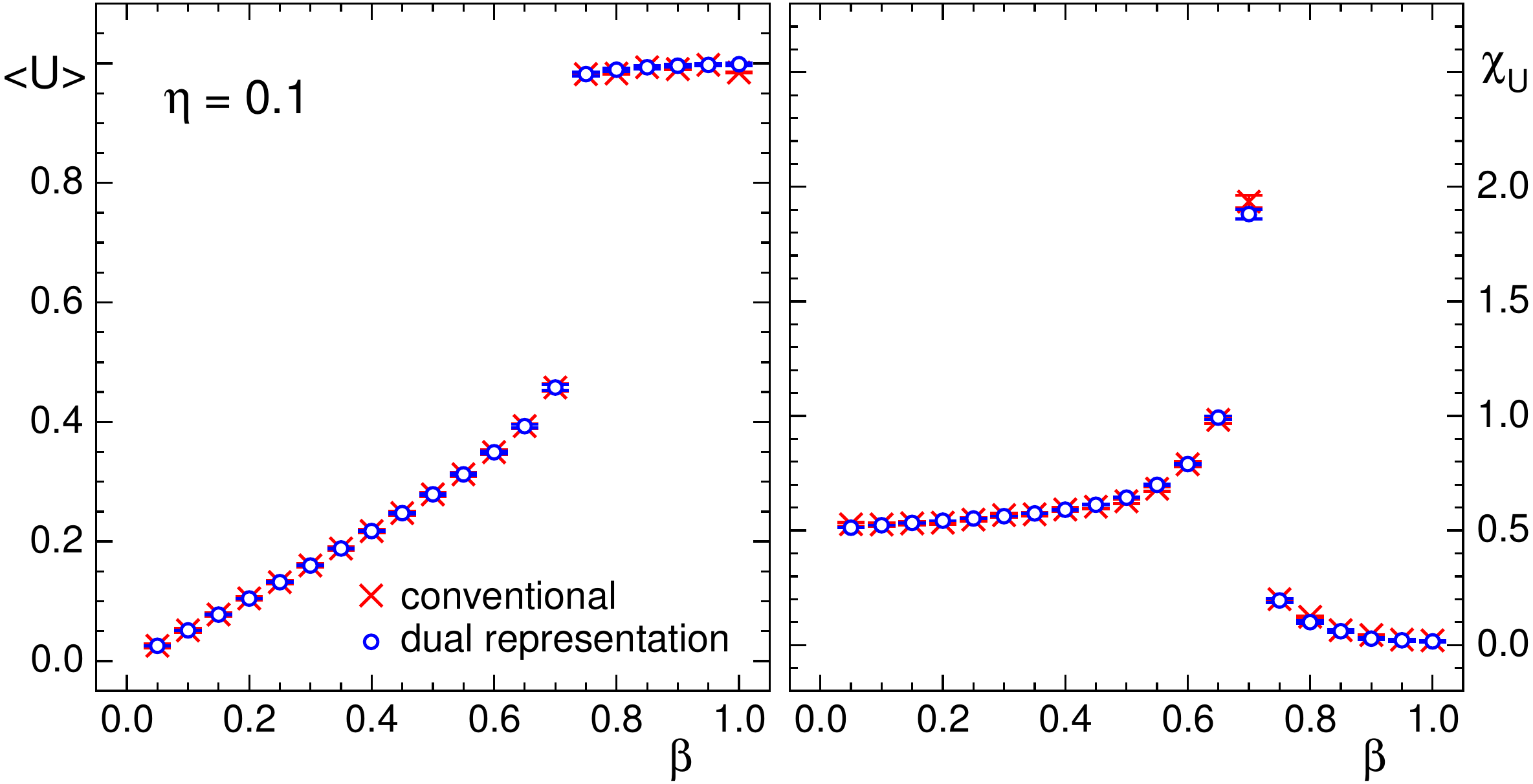}
\caption{Results for $\langle U \rangle$ (lhs.~plot) and $\chi_U$ in full Z$_3$ Gauge-Higgs theory at $\eta = 0.1$ and
$\mu = 0.0$ as a function of the inverse gauge coupling $\beta$. We compare the
conventional (crosses) and the dual approach (circles).}
\label{gh_k01_mu0}
\vskip5mm
\centering
\includegraphics[width=0.75\textwidth,clip]{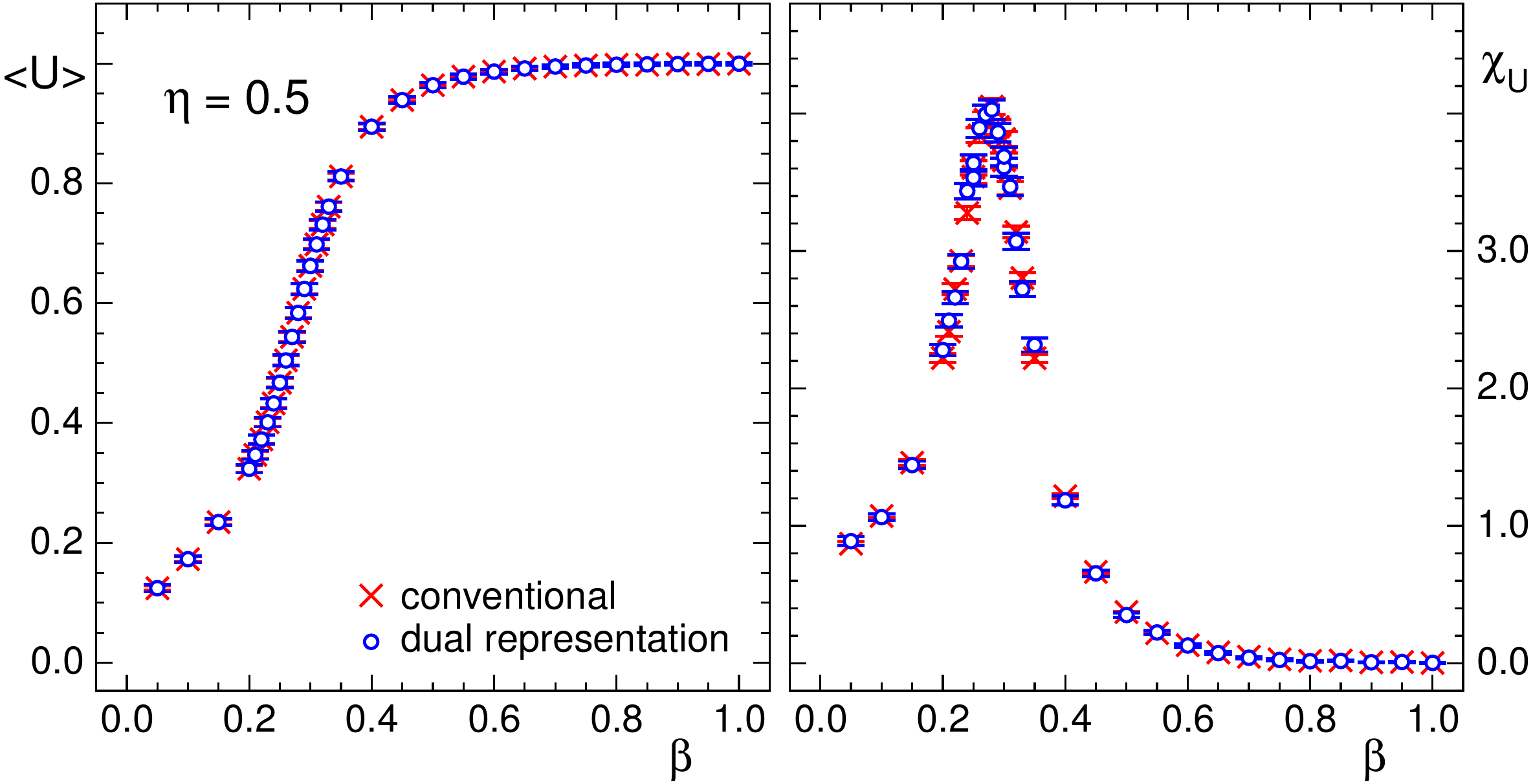}
\caption{Same as in Fig.\protect{\ref{gh_k01_mu0}}, but now for $\eta = 0.5$.}
\label{gh_k05_mu0}
\end{figure*}

At this point we remark, that the case of pure U(1) gauge theory in a dual
representation  which is similar to our Z$_3$ form \footnote{Also for gauge group
U(1) the dual representation of pure gauge theory consists of closed surfaces, but
the plaquette occupation numbers assume values  in the integers $\mathds{Z}$.} 
has been studied  numerically in \cite{u1lgt1,u1lgt2} (see also the discussion in
\cite{reviews2}). The algorithms in \cite{u1lgt1} are very similar to our updates
(however, partly without  the global plane updates). In \cite{u1lgt2} defects
(i.e., boundaries) are introduced for the surfaces and a generalization of the
worm algorithm \cite{worm} to surfaces is explored.

\subsection{The full Z$_3$ Gauge-Higgs model at $\mu = 0$}

Let us now come to the full Z$_3$ Gauge-Higgs model. Again we would like to test
the dual approach and verify its validity by comparison with a conventional
simulation, which is possible for $\mu = 0$.

For checking the correct implementation of the dual approach we compare the
results for simulations at $\mu = 0$ using two different values of the coupling
$\eta$: $\eta = 0.1$ and $\eta = 0.5$ and plot the observables as a function
of $\beta$. In Fig.~\ref{gh_k01_mu0} and Fig.~\ref{gh_k05_mu0} we show the 
results for $\langle U \rangle$ (lhs.\ plots) and $\chi_U$ (rhs.) as a function of
$\beta$. Fig.~\ref{gh_k01_mu0} is for $\eta = 0.1$ and Fig.~\ref{gh_k05_mu0} for
$\eta = 0.5$ and  we compare the results from a conventional simulation
(crosses) to the outcome from the dual approach (circles) and again use volumes 
of size $10^4$ and the same sequence and amount of the different update sweeps
as for pure gauge theory.

For $\eta = 0.1$ the first order transition persists and we see very little
change in our observables $\langle U \rangle$ and $\chi_U$ when comparing the
results to pure gauge theory. Again we observe that the results from the dual
simulation and the conventional approach match very well. However, since at
$\eta = 0.1$ and vanishing chemical potential $\mu$ the influence of the Higgs
field seems to be small, we consider a second, larger value of $\eta$.

The results for $\langle U \rangle$ and $\chi_U$ at $\eta = 0.5$ and $\mu = 0.0$
are shown in Fig.~\ref{gh_k05_mu0}. We now observe quite a change in the behavior
of the observables in comparison to the pure gauge and $\eta = 0.1$ cases. The
phase transition has apparently disappeared and we only find a smooth crossover-type of
behavior between the strong- and weak coupling phases. The maximum of the
susceptibility $\chi_U$ has shifted to rather small values -- the crossover takes
place near $\beta = 0.28$. The important fact is, that also here at a larger value
of $\eta$, where obviously the Higgs field has a much stronger influence, the
results from the conventional approach and the dual simulation agree very well,
again confirming the correctness of the implementation of the dual approach.

\section{The Z$_3$ Gauge-Higgs model at finite density}

Let us now come to the more interesting case of finite density. Here  conventional
simulations fail and the full potential of the dual approach can be unveiled. Before we 
start with the presentation of Monte Carlo results we first discuss some characteristic features
of the dual representation at finite density.

\subsection{Finite density dynamics in the dual representation}

The dual representation of the Z$_3$ Gauge-Higgs model uses two sets of degrees of freedom, the plaquette occupation numbers $p$ and the fluxes $k$. For the analysis of the 
mechanisms that drive the systems at finite density it is useful to think a little bit about the dynamics of the dual variables, and this subsection is devoted to that task. 

The  dual degrees of freedom assume values in $\{-1,0,+1\}$, i.e.,  $p_{x,\sigma \tau} \in \{-1,0,+1\}$ and $k_{x,\nu}  \in \{-1,0,+1\}$.
A trivial value of the plaquette occupation number, i.e., $p_{x,\sigma \tau} = 0$ comes with a Boltzmann factor of $1$ (compare (\ref{Z3gaugeweight})), while 
non-trivial values  $p_{x,\sigma \tau} \pm 1$ give rise to a factor of $B_\beta < 1$ (see (\ref{aux3}) for the definition of $B_\beta$). 
Thus non-trivial values of plaquette occupation numbers $p$ are suppressed 
by their Boltzmann factor. On the other hand configurations with many $p_{x,\sigma \tau} \neq 0$ have a much higher entropy and (as always)  the interplay of entropy and Boltzmann factor gives rise to
the first order transition of the pure gauge theory discussed in Subsection IV-A. The corresponding 
observables $\langle U \rangle$ and $\chi_U$ are simple functions of the plaquette occupation numbers and their fluctuations. 
We stress at this point that both observables have a $\beta$-dependent additive term (compare (\ref{obs1Z3})). For the plaquette expectation value the additive term 
is given by $B_\beta$, and   $\langle U \rangle$ is non-vanishing for $\beta > 0$ even when all plaquette occupation numbers $p$ are trivial, since $B_\beta > 0$ for $\beta > 0$.

Similar to the plaquette occupation numbers, the spatial flux variables $k_{x,j}, j = 1,2,3$, 
have a Boltzmann factor of 1 for $k_{x,j} = 0$ and a Boltzmann factor $B_\eta < 1$ for $k_{x,j} = \pm 1$ (see (\ref{Z3gaugeweight})).
As for the case of the plaquette occupation numbers, we find for the $k$-variables that trivial values of the spatial fluxes are preferred by the Boltzmann factor. The temporal 
flux variables $k_{x,4}$ are connected with the Boltzmann factors $M_s$ with $s \in \{ -1,0,1\}$ defined in (\ref{aux2}). For $\mu > 0$ we have $M_{+1} > M_{-1}$ (see also the discussion below) and 
temporal flux with $k_{x,4} = +1$ is favored over negative temporal flux, i.e., $k_{x,4} = -1$.

\begin{figure}[b!]
\centering
\hspace*{-1.5mm}
\includegraphics[width=0.48\textwidth,clip]{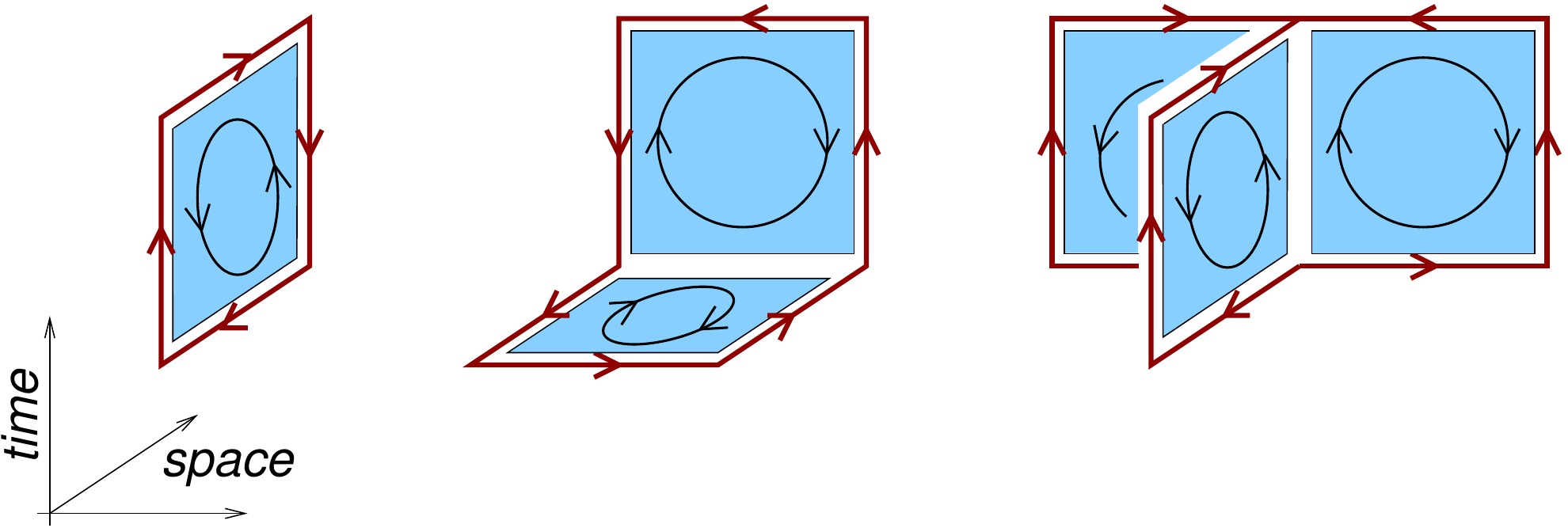}
\caption{Examples of low-lying excitations in the dual representation of the Z$_3$ Gauge-Higgs model. We use red thick lines with arrows for the $k$ flux variables and 
blue squares (with a circle showing the orientation) for the plaquette occupation numbers $p$. The rules for admissible configurations dictate 
that at each site the total flux from the $k$-variables has to be a multiple of 3. In addition for each 
link the combined flux of $k$-variables and plaquette occupation numbers $p$ also has to be a multiple of 3. }
\label{excitations}
\end{figure}

To illustrate the physical picture in terms of the dual representation, in Fig.~\ref{excitations} we show a few low-lying excitations of the
Z$_3$ Gauge-Higgs model in the dual representation. Thick red lines oriented with arrows are used for the $k$-flux and filled blue squares for
non-vanishing plaquette occupation numbers, and the circles in the squares indicate the orientation of the plaquette according to the sign 
of the corresponding plaquette occupation number $p_{x,\sigma \tau}$.  
The simplest excitations (the lhs.\ diagram in Fig.~\ref{excitations}) are 
an occupied plaquette surrounded by flux.  At each link the flux is compensated by the plaquette. Occupied plaquettes 
with suitable relative orientation can be attached to each other (see the example in the center diagram). At the link where they are joined the 
flux is absent and the contributions from the plaquette occupation numbers compensate. 
Finally the excitation in the 
rhs.\ diagram makes use of the fact that flux and plaquette numbers need to vanish only modulo 3. Three units of flux
emerge from a site, travel in time and then terminate at another site. The constraints are again saturated with plaquettes,
such that at the central temporal link three plaquettes together obey the constraint. This excitation, which resembles a baryon, 
carries three factors of $M_{+1}$ and thus is enhanced by the chemical potential.  

Without chemical potential the Boltzmann factor for the excitations shown in Fig.~\ref{excitations} follows a simple rule: The more non-vanishing
plaquette occupation numbers $p$ and fluxes $k$ a configuration has, the lower is the corresponding Boltzmann weight, and the dynamics discussed 
in Subsection IV-B.\ for the full Z$_3$ Gauge Higgs system at $\mu = 0$ is again determined by the interplay of entropy and Boltzmann weight.

\begin{figure*}[t!]
\centering
\includegraphics[width=0.75\textwidth,clip]{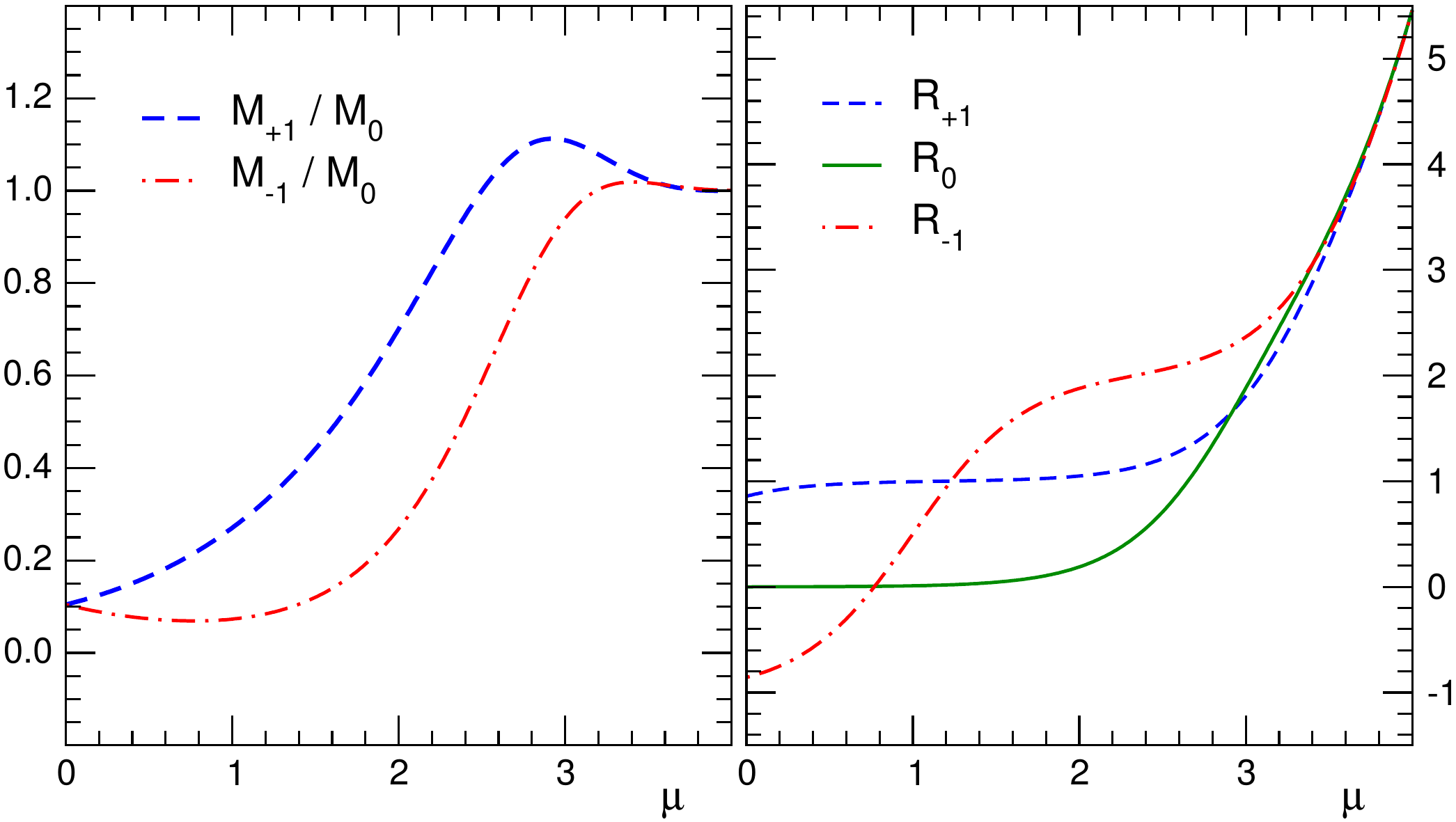}
\caption{Lhs.: The ratios $M_{+1}/M_{0}$ and $M_{-1}/M_0$ that determine the transition from temporal flux $k_{x,4} = 0$ to $k_{x,4} = +1$ and  $k_{x,4} = -1$
as a function of $\mu$ at $\eta = 0.1$.  Rhs.: The coefficients $R_s = M_{s}^\prime /M_s$ that determine the contributions of temporal flux $k_{x,4} = s$ with $s \in 
\{-1,0,+1\}$ to the particle number density $n$ in the dual representation (\ref{obs2Z3}).}
\label{muplots}
\end{figure*}

The situation is more complex when the chemical potential $\mu$ is turned on. Then the weights for the temporal flux variables  $k_{x,4}$ obey $M_{+1} > M_{-1}$ and the probability for
positive temporal fluxes is increased relative to the probability for negative fluxes. In the lhs.\ plots of Fig.~\ref{muplots} we show 
the ratios $M_{+1}/M_0$ and $M_{-1}/M_0$ as a function of $\mu$ 
using $\eta = 0.1$. The Metropolis probabilities for accepting a step from $k_{x,4} = 0$ to $k_{x,4} = +1$ and $k_{x,4} = -1$, respectively are given by
$\max(1, M_{+1}/M_0)$ and $\max(1, M_{-1}/M_0)$. We see that for
all $\mu \neq 0$ we have $M_{+1}/M_0 > M_{-1}/M_0$ and temporal fluxes $k_{x,4} = +1$ are always favored over negative ones. For values of $\mu$ up to $\mu \sim 2.8$ the discrepancy 
between  $M_{+1}/M_0$ and $M_{-1}/M_0$ remains large and the non-zero chemical potential pumps positive temporal $k$-flux into the system. However, this $k$-flux has to be compensated by
plaquettes which costs Boltzmann weight and dampens the increase of  temporal flux with $k_{x,4} = +1$. This interplay between the pumping with $\mu$ and the damping by the
Boltzmann factor of the plaquette variables can give rise to a phase transition at some critical value of $\mu$, when the positive temporal $k$-flux starts to dominate 
and drags along the plaquette occupation numbers. In the next subsection we will see that this is indeed the case for suitable values of the couplings.

It is, however, important to note that for $\mu > 2.8$ the
two ratios  $M_{+1}/M_0$ and $M_{-1}/M_0$ start to approach each other again and they both have limit 1 for  large $\mu$. This implies that at large values of 
$\mu$ the probabilities for temporal fluxes $k_{x,4}=+1$, $k_{x,4}=0$
and $k_{x,4}=-1$ are equal, and the chemical potential does not favor an orientation for the $k_{x,4}$. Thus for sufficiently large $\mu$ the
Boltzmann factor of the plaquette occupation numbers may again dominate the physics. This raises an interesting question that 
then has to be answered in this region: How can the particle number density $n$ keep growing with $\mu$ if the weights $M_s$ that drive the $\mu$-dependence for small $\mu$ 
become degenerate at large $\mu$? The answer lies in the flux representation of
$n$ given in (\ref{obs2Z3}): The numbers ${\cal N}_s, s \in \{+1,0,-1\}$, for temporal links with $k_{x,4} = s$ are weighted with the factors $R_s = M_s^\prime / M_s$. 
Also these ratios approach each other for large $\mu$ and all three grow as $e^{\mu}$  (rhs.\ plot in Fig.~\ref{muplots}). 
Using ${\cal N}_{+1} +{\cal N}_{0} +{\cal N}_{-1}  = N_s^3 N_t$ we conclude from
(\ref{obs2Z3}) that 
for large $\mu$ we have $n \propto R_s \propto \eta e^\mu$. In the plots for $n$ and $\chi_n$ of Fig.~8 we display these limiting curves and find that the Monte Carlo data nicely approach 
the expected asymptotic behavior. We found similarly good asymptotic behavior also for the weak coupling results shown in Fig.~11, but since there we use a smaller interval on the x-axis this is not entirely obvious from the plots. 

It is a remarkable feature of the dual representation that part of the behavior of observables (e.g., the asymptotic behavior in the above example)
is already encoded in the expansion factors of the partition sum 
and the representation of the observables in terms of dual variables.

\subsection{Finite density results at strong coupling}

Let us now come to the numerical results at finite density.
We numerically analyzed the finite density behavior for both values of $\eta$ that were
considered in Section IV, i.e., $\eta = 0.1$ and $\eta = 0.5$. For
the $\eta = 0.5$ case we did not find transitions as a function of $\mu$ and thus
do not present the corresponding results in this exploratory study. Instead  we
focus on $\eta = 0.1$ and study the system for two couplings  on both sides of
the transition located at $\beta_c \sim 0.7$. We begin with the value $\beta = 0.6$ in the strong
coupling region in this subsection, continuing with $\beta = 0.8$ in the weak coupling regime
in the next subsection. 
 
For the
finite density study we use lattices of sizes $N_s^3 \times 50$, with $N_s$ ranging
from 2 to 12. The reason for the large $N_t = 50$ is that we want to study the
system at zero temperature. Although  with $N_t = 50$ the temperature $T = 1/50 =
0.02$ in lattice units is not exactly zero, this value is much smaller than any
other involved scale and constitutes a good approximation of $T = 0$. 
We  consider
the full set of observables discussed in Section III, i.e., in addition to 
the plaquette $\langle U \rangle$ and its susceptibility $\chi_U$
we now also analyze the particle number density $n$ and its
susceptibility $\chi_n$.

\begin{figure*}[t]
\centering
\includegraphics[width=0.75\textwidth,clip]{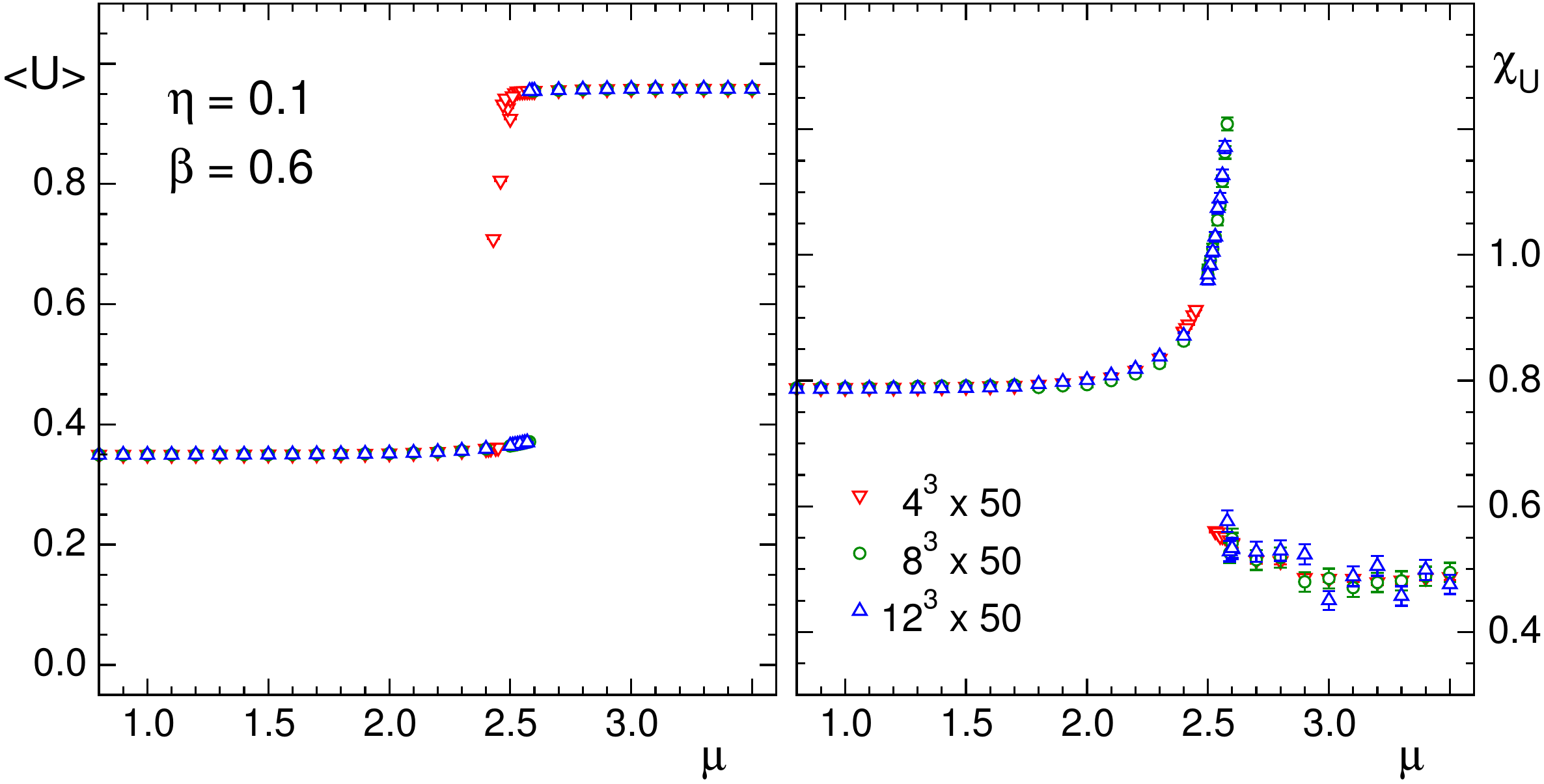}
\vskip5mm
\hspace*{1.3mm}
\includegraphics[width=0.743\textwidth,clip]{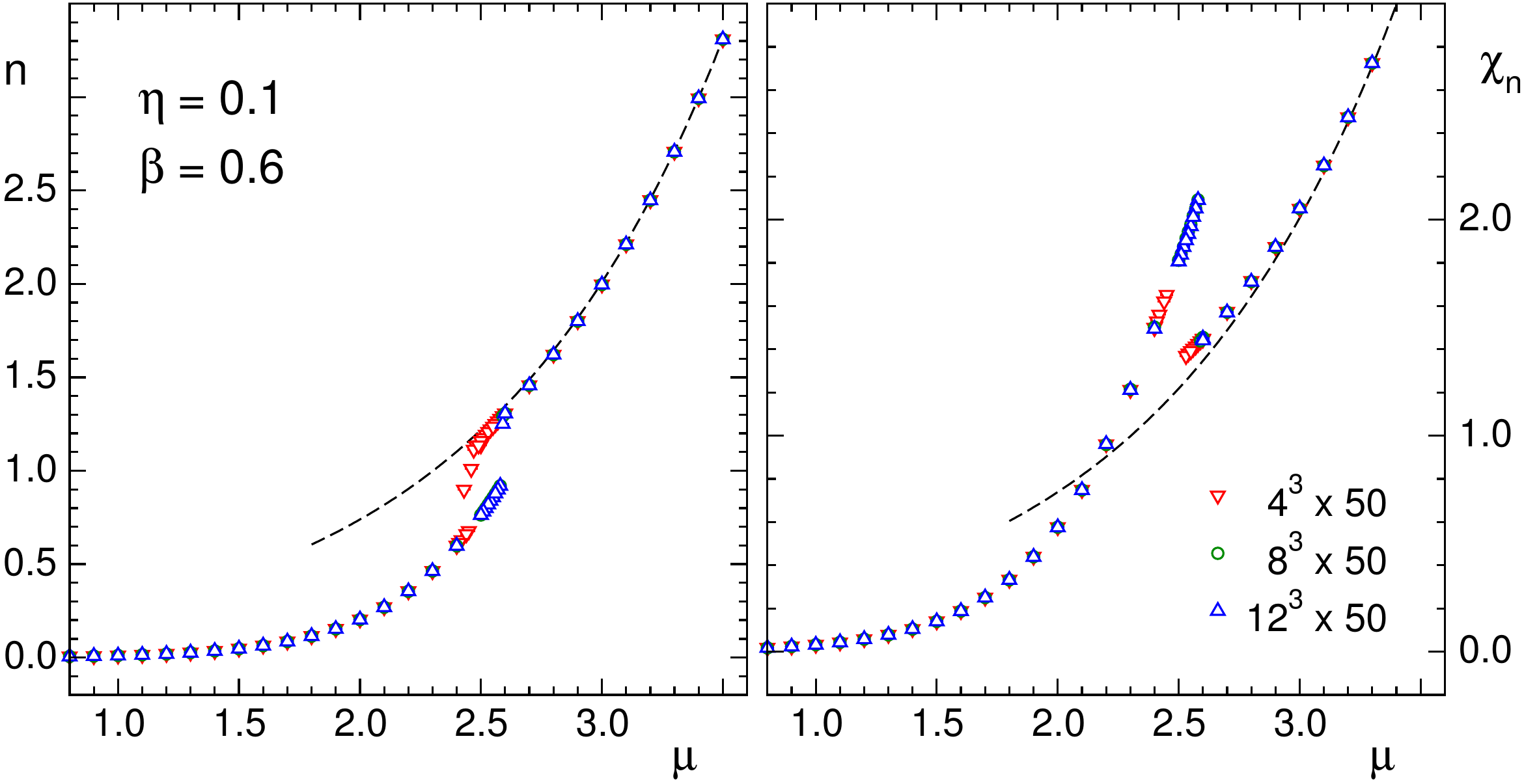}
\caption{Results for $\langle U \rangle$, $\chi_U$,
$n$ and $\chi_n$ in the strong coupling region of 
Z$_3$ Gauge-Higgs theory ($\eta = 0.1$ and
$\beta = 0.6$) as a function of the chemical potential $\mu$. 
We compare the results for three different spatial volumes with $N_s = 4$, 8 and 12. The dashed curves
in the plots for $n$ and $\chi_n$ are the asymptotic curves expected from the behavior of the 
coefficients $M_s$ for large $\mu$ (see the discussion in Subsection V-A.).
\label{gh_k01_b06_allmu}}
\end{figure*}

In Fig.~\ref{gh_k01_b06_allmu} we show the results for the four observables as a
function of the chemical potential $\mu$ and compare runs on $N_s^3 \times 50$ 
lattices with different spatial extents $N_s = 4$, 8 and 12. The data for $N_s = 8$
and $N_s = 12$ fall on top of each other and only the $N_s = 4$ data show a slight
discrepancy near $\mu \sim 2.57$ due to finite volume effects (the transition zone is 
shifted towards a slightly smaller $\mu$). It is obvious that for
$\mu = \mu_c \sim 2.57$ the system undergoes a first order transition: Both first
derivatives, the plaquette expectation value $\langle U \rangle$ and the particle
number density $n$ show a clear discontinuity. The corresponding susceptibilities 
$\chi_U$ and $\chi_n$ diverge at $\mu_c$, which is, however, somewhat hard to see
here since the transition is so sharp: As a matter of fact for very finely spaced
values of $\mu$ near $\mu_c$ we find values for $\chi_n$ that are three orders of 
magnitude larger than the scale used in the plot. A second indication that the
transition is very narrow is the fact that only for $N_s = 4$ we see small finite
volume effects: The transition is shifted slightly to the left and appears somewhat
rounded (at least for the first derivatives $\langle U \rangle$ and $n$). This trend towards visible finite size effects 
continues when using $N_s = 2$ (data not shown). 

Finally, a dataset
at  $12^3 \times 100$, which corresponds to an even lower temperature of $T = 0.01$
in lattice units, falls on top of the $12^3 \times 50$ data.  We conclude that we
reliably describe the situation at zero temperature, and that for $\eta = 0.1$ and
$\beta = 0.6$ the system undergoes a very narrow first order transition at $\mu_c
\sim 2.57$.

From the fact that the plaquette undergoes such a drastic change at $\mu_c$ we conclude that the gauge 
dynamics plays an important role in the transition. This is also reflected in the finding that 
the critical value $\mu_c$ depends on $\beta$. For $\beta = 0.6$ we observed $\mu_c \sim
2.57$, while for $\beta = 0.65$ it is at $\mu_c \sim 2.35$, and $\mu_c \sim 3.0$ for 
$\beta = 0.55$. 
  
\begin{figure*}[t]
\underline{$\beta = 0.6, \mu = 2.4$:}
\vskip3mm
\includegraphics[width=0.85\textwidth,clip]{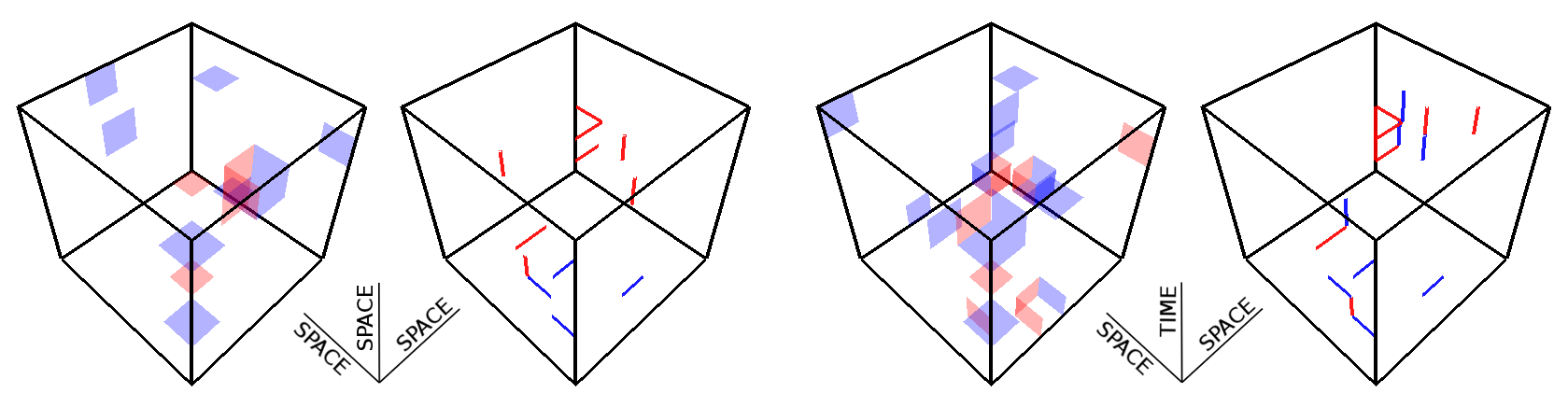}
\vskip5mm
\underline{$\beta = 0.6, \mu = 2.7$:}
\vskip3mm
\includegraphics[width=0.85\textwidth,clip]{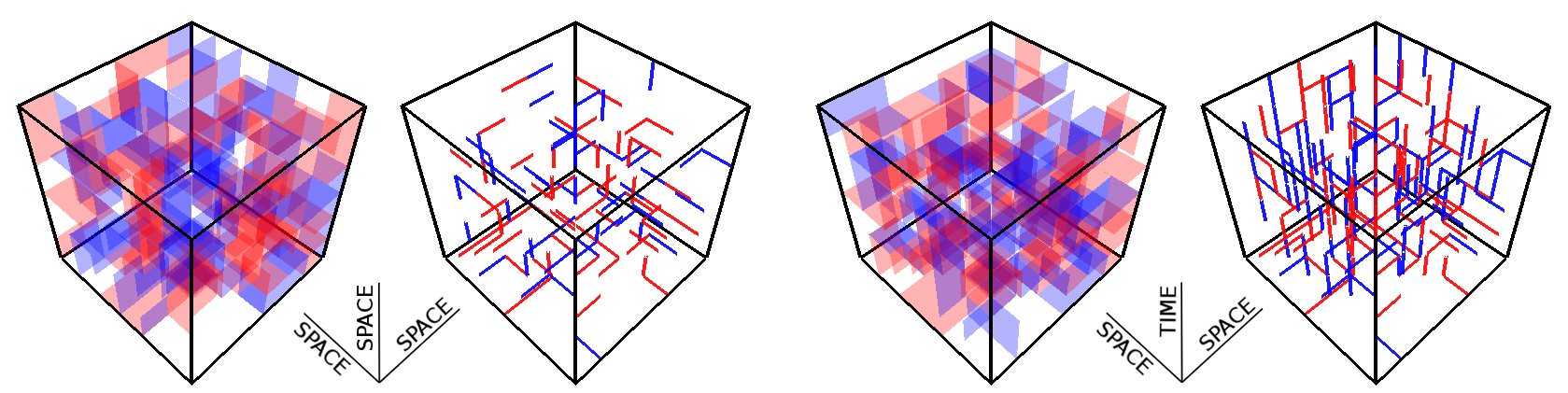}
\vskip5mm
\caption{3-D illustration of typical configurations of plaquette occupation numbers $p$ and fluxes $k$ in the strong coupling phase ($\beta = 0.6, \eta = 0.1$)
for $\mu = 2.4  < \mu_c$ (top row of plots) and for $\mu = 2.7 > \mu_c$ (bottom). We use 3-dimensional sections of the lattice embedded in 4 dimensions
and show purely spatial sections (1st and 2nd plot in each row) and sections where the vertical direction is time (3rd and 4th plot). In each pair the lhs.\ section
shows the non-trivial plaquette occupation numbers and we use blue for plaquettes with $p = +1$ and red for $p = -1$. Likewise, in the rhs.\  plot of each pair 
we show the non-trivial link variables with blue links for $k = 1$ and red for $k = -1$. }
\label{b06_3d}
\end{figure*}

Finally we point out that for large $\mu$ the results for $n$ and $\chi_n$ approach the asymptotic curves $\propto \eta \exp(\mu)$ 
(dashed lines in the plots) which we derived in
Subsection V-A.\ from the
behavior of the coefficients $M_s$.
 
To further understand the nature of the transition, in Fig.~\ref{b06_3d} we show 3-D illustrations  
of the plaquette and flux occupation numbers for a value of $\mu = 2.4 < \mu_c$ (top row of plots) 
and $\mu = 2.7 > \mu_c$. We consider 3-dimensional sections of the 4-dimensional lattice with only spatial directions (lhs.\ pair of plots in each row)
and 3-dimensional sections with the vertical direction being time (rhs.\ pair of plots). In each pair the
lhs.\ section illustrates the plaquette occupation numbers (plaquettes with $p = +1$ are blue, those with
$p = -1$ are red), while the rhs.\ section of each pair displays the corresponding fluxes (links with $k = +1$ are blue, those with 
$k = -1$ are red). 

It is obvious that below $\mu_c$ (top row of plots) the occupation numbers for plaquettes and for fluxes
are very small, while for $\mu > \mu_c$ (bottom) we see a large abundance of non-zero occupation numbers. The transition 
thus is between a dilute phase where all occupation numbers are small and a condensed phase characterized by high occupation
numbers for both, fluxes and plaquettes. Furthermore, a close inspection of the 3-D plot for the fluxes at $\mu > \mu_c$ on the temporal section 
(lower right plot) shows that positive temporal flux (vertical blue lines in the plot) dominates, as expected from the finite density picture in terms
of the dual variables discussed in the previous subsection.  

We remark at this point that the value of $\langle U \rangle \sim 0.35$ which we observe
below $\mu_c$ is mainly due to the constant term in the dual representation (\ref{obs1Z3}). Here we use $\beta = 0.6$ for the inverse 
gauge coupling and the constant term has a value of 0.327. Thus also at very low plaquette occupation numbers $p$ the plaquette 
has a value of  $\langle U \rangle \sim 0.35$ at $\beta = 0.6$. The jump of $\langle U \rangle \sim$ from 0.35 
to almost 1 at $\mu_c$ is due to the condensation of non-trivial 
plaquette occupation numbers $p$ as is obvious from Fig.~\ref{b06_3d}. 

\begin{figure}[t!]
\centering
\hspace*{-1mm}
\includegraphics[width=0.48\textwidth,clip]{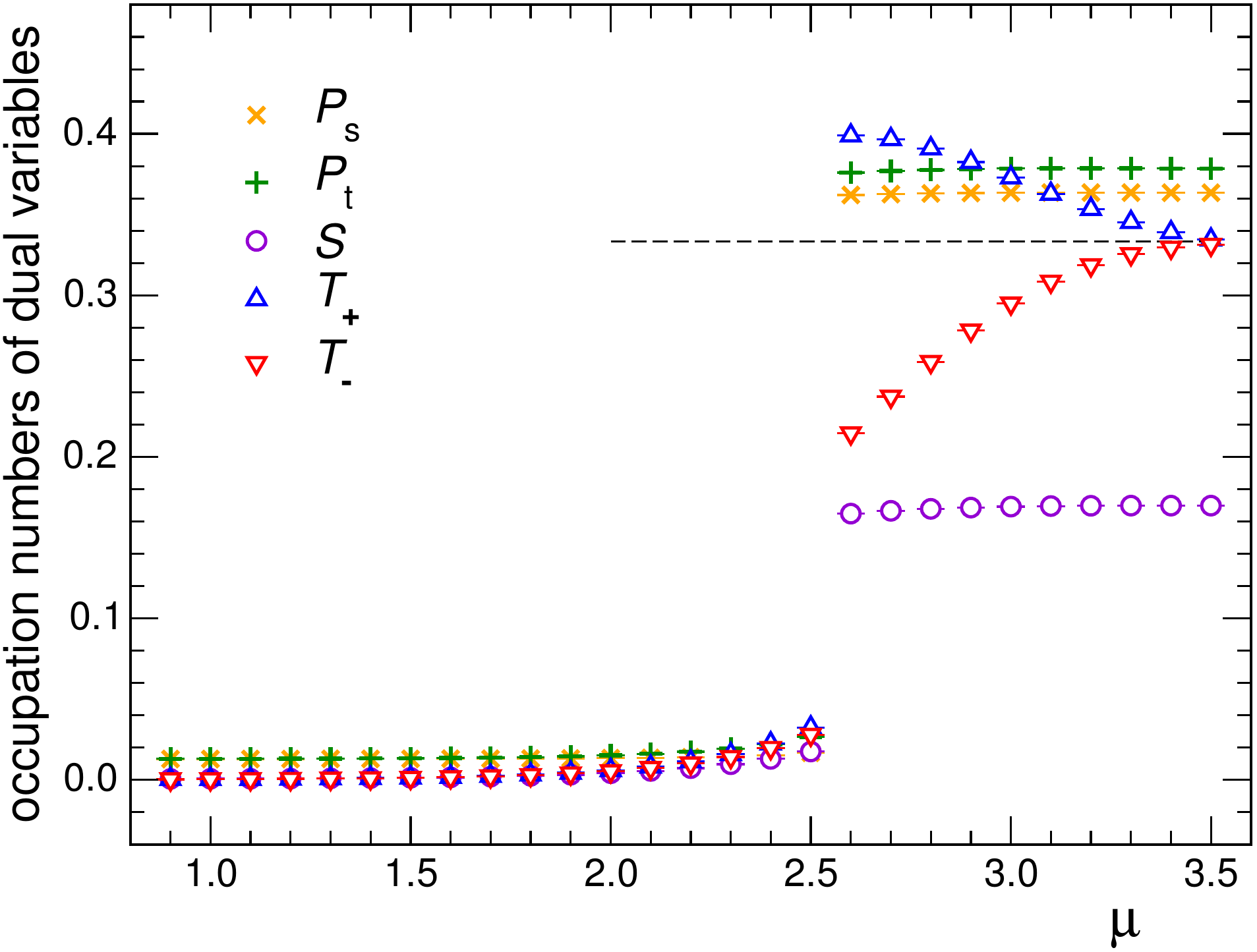}
\caption{Occupation numbers for the dual variables at strong coupling ($\beta = 0.6, \eta = 0.1$) as a function of $\mu$ from our $12^4 \times 50$ lattices. 
We show the number of non-trivial spatial plaquettes $P_s$, i.e., the total number of $p_{x,\sigma\tau} \neq 0$, the number of temporal 
non-trivial plaquettes $P_t$ ($p_{x,\sigma 4} \neq 0$), the number of non-trivial spatial fluxes $S$ ($k_{x,j} \neq 0$), and the numbers of positive and negative
temporal fluxes $T_+$ and $T_-$ ($k_{x,4} = +1$ and $k_{x,4} = -1$, respectively).
All occupation numbers are given as intensive quantities, normalized such that the maximally
possible occupation number is 1. The horizontal line marks the value 1/3 which the occupation numbers $T_+$ and $T_-$ are expected to approach for large $\mu$.}
\label{occupation06}
\end{figure}

To further illuminate the picture of a condensation of the plaquette occupation numbers triggered by temporal $k$-flux enhanced by $\mu$,
we now look at plots for the occupation numbers for plaquettes and fluxes as a function of the chemical potential $\mu$. Fig.~\ref{occupation06}
shows the number of non-trivial spatial plaquettes $P_s$, i.e., the total number of $p_{x,\sigma\tau} \neq 0$, the number of temporal 
non-trivial plaquettes $P_t$ ($p_{x,\sigma 4} \neq 0$), the number of non-trivial spatial fluxes $S$ ($k_{x,j} \neq 0$), and the numbers of positive and negative
temporal fluxes $T_+$ and $T_-$ ($k_{x,4} = +1$ and $k_{x,4} = -1$). All these occupation numbers are normalized such that the maximally possible occupation
number is 1. 

The occupation numbers for spatial ($P_s$) and temporal ($P_t$) plaquettes essentially vanish below $\mu_c$ and then take a jump where both 
show an occupation of roughly 0.37, with the temporal plaquettes slightly enhanced. This behavior clearly reflects the condensation of the plaquettes we had
discussed above. Similarly the occupation number $S$ for the spatial flux nearly vanishes below $\mu_c$ where it jumps to a finite value. The occupation numbers 
for positive and negative temporal flux,  $T_+$ and $T_-$ show an interesting behavior: Both essentially vanish below $\mu_c$. At the transition their degeneracy is
lifted and both jump to different values. As the chemical potential is increased further the occupation numbers for positive and negative temporal flux  
both approach the value 1/3 (marked by a dashed horizontal line) which is the value one expects when the corresponding Boltzmann weights $M_s$ become degenerate 
(see the discussion in the previous subsection where we analyzed the  behavior of the weight factors $M_s$  shown in Fig.~\ref{muplots}).

\subsection{Finite density results at weak coupling}

\begin{figure*}[t]
\centering
\includegraphics[width=0.76\textwidth,clip]{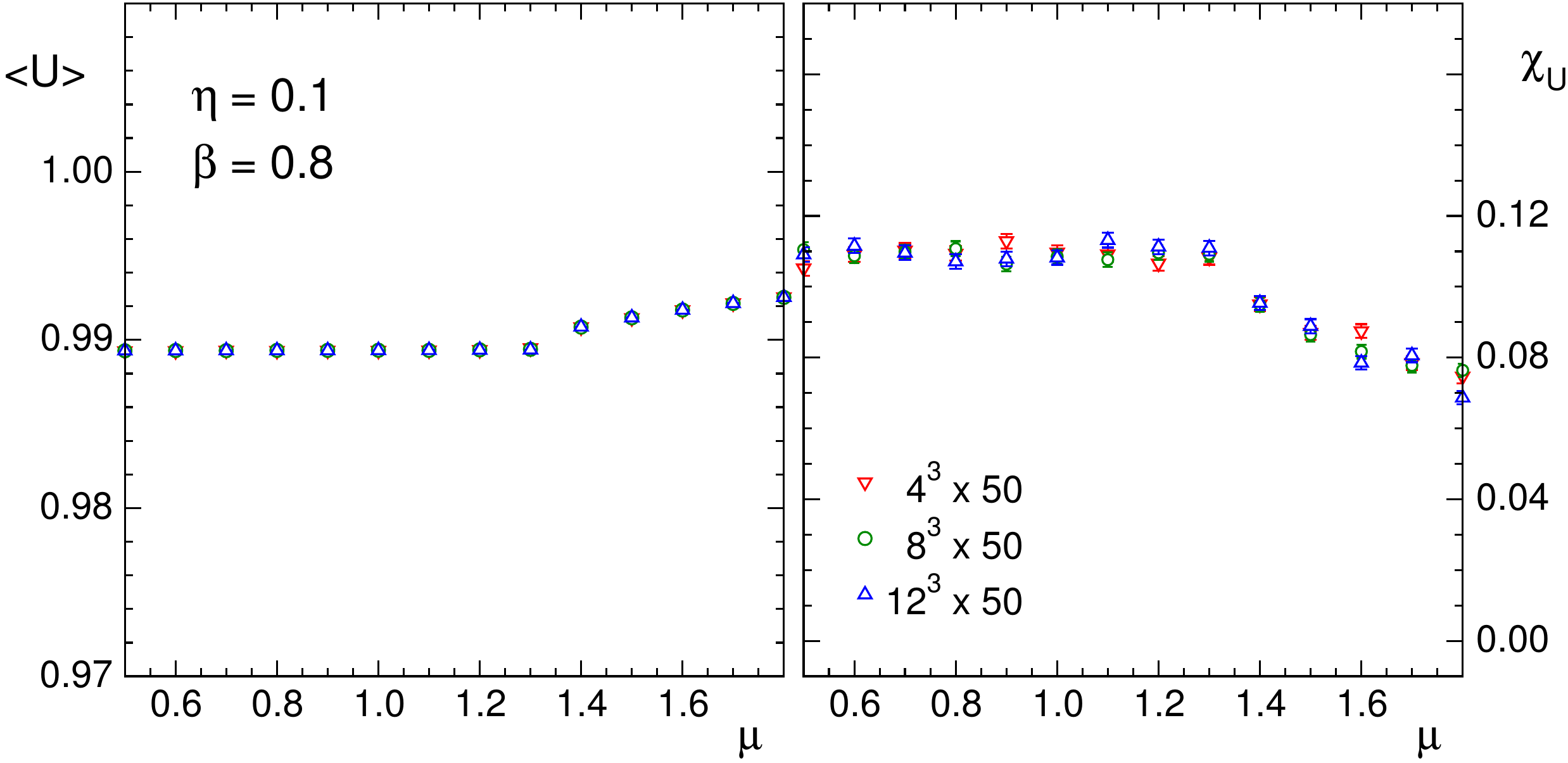}
\vskip5mm
\hspace*{1.2mm}
\includegraphics[width=0.723\textwidth,clip]{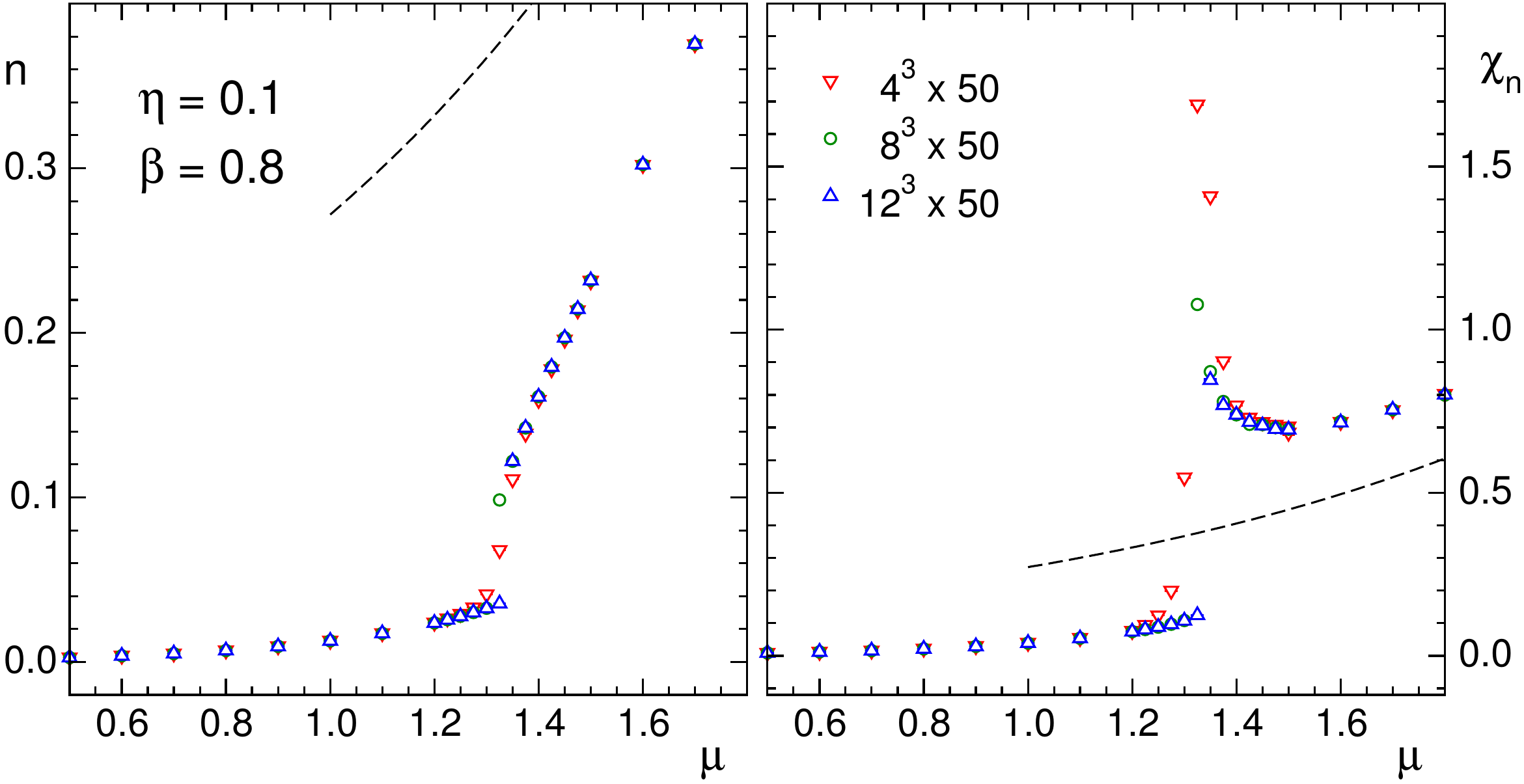}
\caption{Results for $\langle U \rangle$, $\chi_U$,
$n$ and $\chi_n$ in the weak coupling regime of  Z$_3$ Gauge-Higgs theory ($\eta = 0.1$ and
$\beta = 0.8$) as a function of the chemical potential $\mu$. 
We compare the results for three different spatial volumes with $N_s = 4$, 8 and 12.
The dashed curves in the bottom plots represent the asymptotic behavior of $n$ and $\chi_n$ at large $\mu$.}
\label{gh_k01_b08_allmu}
\end{figure*}

We continue with the discussion of the finite density behavior for the weak coupling
region, i.e., for $\beta = 0.8$ and again $\eta = 0.1$. Here the situation is different since
in the weak coupling phase already at $\mu = 0$ we have $\langle U \rangle \sim 1$, i.e., the plaquette occupation numbers
are already condensed. As in the previous 
subsection we consider the plaquette $\langle U \rangle$, its susceptibility $\chi_U$,
the particle number density $n$ and the corresponding 
susceptibility $\chi_n$. In Fig.~\ref{gh_k01_b08_allmu} we show our results as a function of $\mu$,
 comparing runs on $N_s^3 \times 50$ 
lattices with different spatial extents $N_s = 4$, 8 and 12. 
 
The susceptibility $\chi_n$ shows clearly that also in the weak coupling regime at $\beta = 0.8$ we find a 
phase transition which is located at $\mu_c \sim 1.35$. Inspection of the particle number density
shows that $n$ has a discontinuity at  $\mu_c$ and we are again dealing with a first order transition 
here. As for the transition in the strong coupling regime which we discussed in the last subsection, 
also here the finite volume effects are rather small and are visible only for the $N_s = 4$ data.
The transition is very narrow and as before we remark that the maxima in our data for $\chi_n$ are much higher than the
range used for the vertical axis. 

For the plaquette expectation value $\langle U \rangle$ the situation is different from the strong coupling phase
of the last subsection. Since here we start with $\langle U \rangle \sim 1$ below $\mu_c$, the change of
$\langle U \rangle$ at $\mu_c$ is rather unspectacular with the plaquette simply developing a mild   
slope. Correspondingly the susceptibility $\chi_U$ shows a slight drop above $\mu_c$ as the plaquette
numbers become completely saturated. 
We conclude that the transition in the weak coupling phase is predominantly driven by the matter fields,
i.e., the flux variables $k$ in the dual language. This finding is supported  by the fact that changing the coupling $\beta$ from
0.8 to nearby values in the weak coupling regime ($\beta = 0.75$ and $\beta = 0.85$) has no noticeable
effect on the value of $\mu_c$. This underlines the statement that the transition is driven by the flux 
variables in the background of condensed plaquette occupation numbers $p$.  

Again we also compare the results for $n$ and $\chi_n$ to the expected asymptotic behavior (dashed curves in the plot). We found that as in the strong coupling case 
the lattice data very nicely approach the expected asymptotic behavior (although this is not very clearly visible for the range of $\mu$-values chosen in Fig.~\ref{gh_k01_b08_allmu}).

\begin{figure*}[t]
\underline{$\beta = 0.8, \mu = 1.3$:}
\vskip3mm
\hspace*{-3.5mm}
\includegraphics[width=0.85\textwidth,clip]{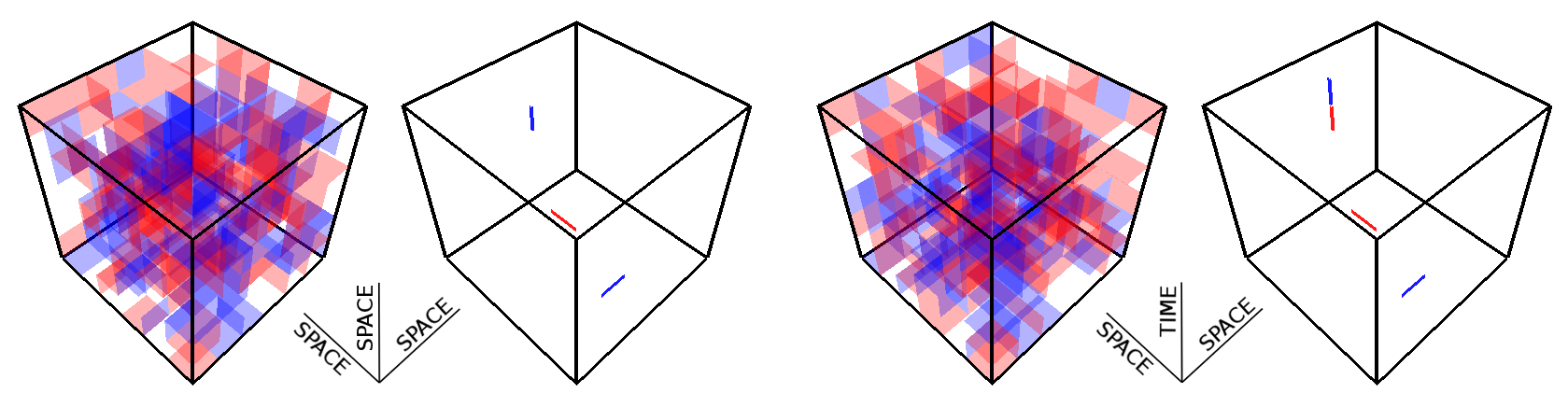}
\vskip5mm
\underline{$\beta = 0.8, \mu = 1.5$:}
\vskip3mm
\hspace*{-3.5mm}
\includegraphics[width=0.85\textwidth,clip]{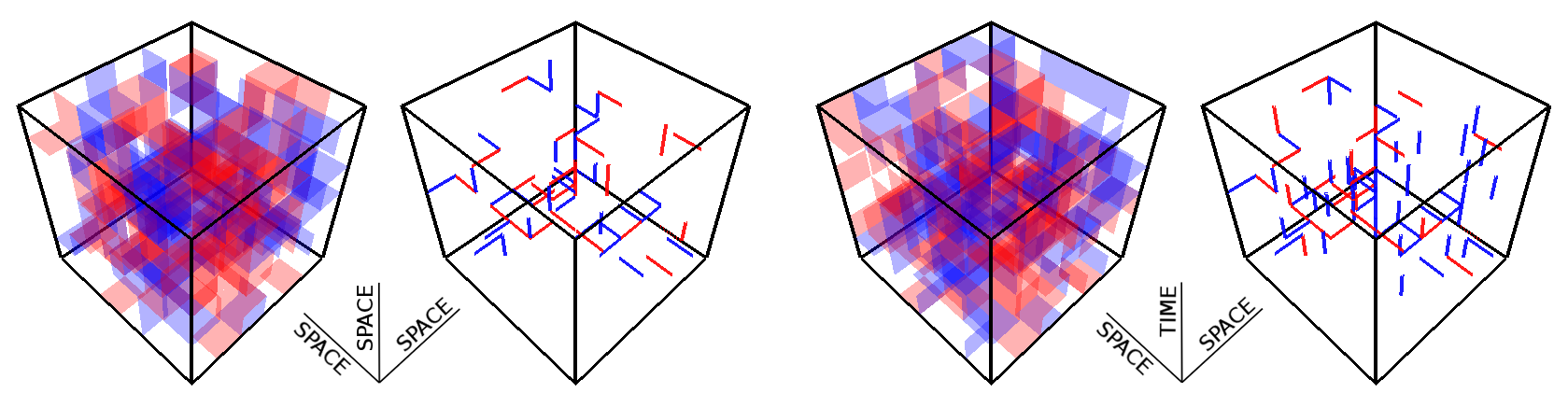}
\vskip5mm
\caption{3-D illustration of typical configurations of plaquette occupation numbers $p$ and fluxes $k$ in the weak coupling phase ($\beta = 0.8, \eta = 0.1$)
for $\mu = 1.3  < \mu_c$ (top row of plots) and for $\mu = 1.5 > \mu_c$ (bottom). We use 3-dimensional sections through the lattice embedded in 4 dimensions
and show purely spatial sections (1st and 2nd plot in each row) and sections where the vertical direction is time (3rd and 4th plot). In each pair the lhs.\ plot
shows the non-trivial plaquette occupation numbers and we use blue for plaquettes with $p = +1$ and red for $p = -1$. Likewise, in the rhs.\ plot of each pair 
we show the non-trivial link variables with blue links for $k = 1$ and red for $k = -1$. }
\label{b08_3d}
\end{figure*}

In order to analyze the nature of the transition in the weak coupling regime, in Fig.~\ref{b08_3d} we repeat the 3-D illustration of
the plaquette occupation numbers and fluxes of the previous subsection. It is obvious that the plaquette occupation numbers are large on 
both sides of the transition, and the fluxes undergo their transition in a condensed medium of plaquette occupation numbers. The flux transition is again manifest in
an abrupt increase of flux, although with a smaller amplitude than in the strong coupling phase. As before we see a dominance of positive
temporal flux above $\mu_c$, i.e, vertical blue lines in the lower right plot.

\begin{figure}[h!]
\centering
\hspace*{-1mm}
\includegraphics[width=0.48\textwidth,clip]{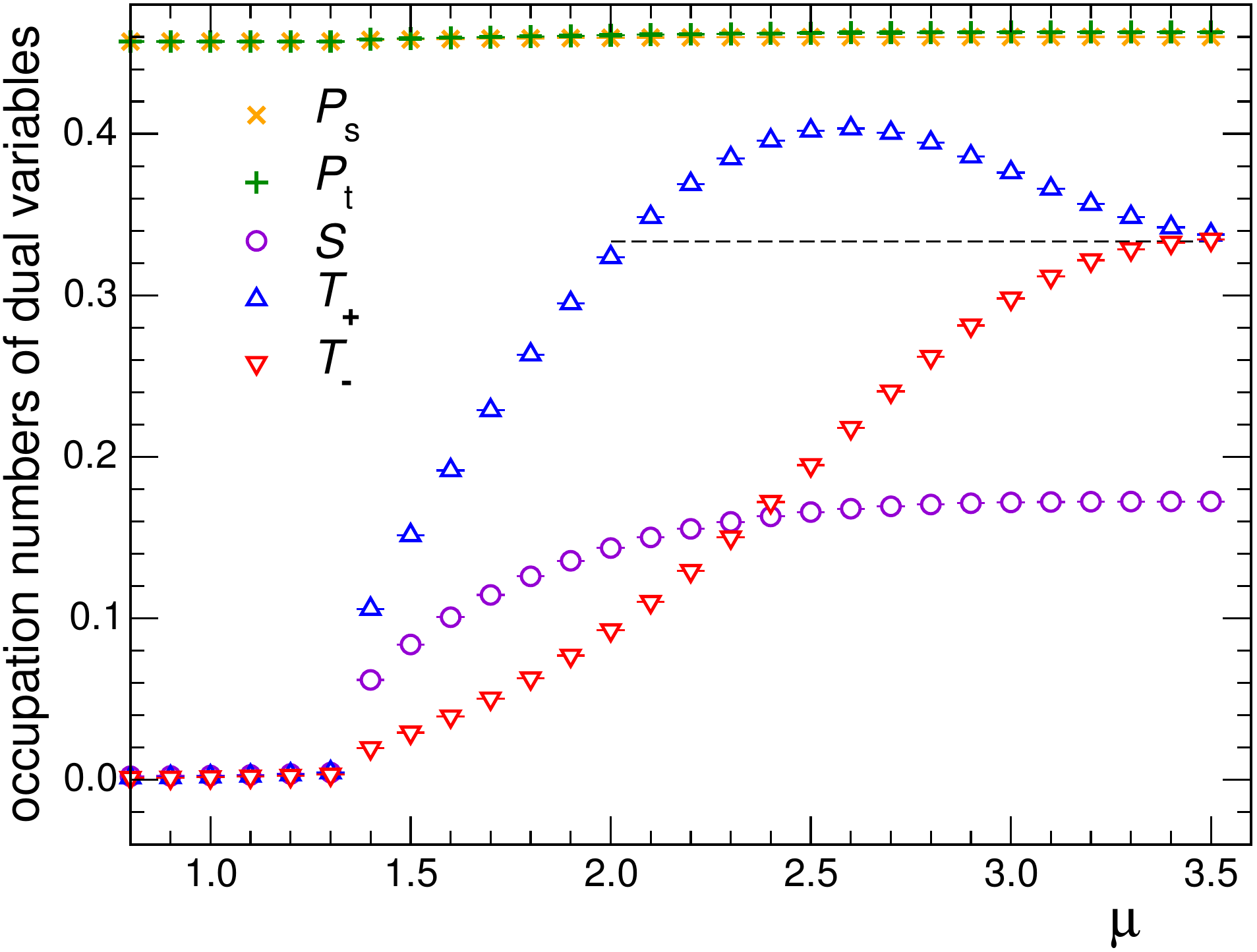}
\caption{Same as Fig.~11 now at $\beta = 0.8, \eta = 0.1$: 
We show the number of non-trivial spatial plaquettes $P_s$, the number of temporal 
non-trivial plaquettes $P_t$, the number of non-trivial spatial fluxes $S$ and the numbers of positive and negative
temporal fluxes $T_+$ and $T_-$.
The horizontal line marks the asymptotic value 1/3 for $T_+$ and $T_-$.}
\label{occupation08}
\end{figure}

We conclude  the discussion of the transition in terms of the dual variables by again analyzing the occupation numbers as a function of the chemical potential.
In Fig.~\ref{occupation08}  we show the weak coupling results for non-trivial occupation numbers of spatial and temporal plaquettes ($P_s, P_t$), of spatial flux (S), 
and the occupation numbers for positive and negative temporal flux ($T_+, T_-$). We use the same definitions as in Fig.~\ref{occupation06} for the strong coupling phase.
Obviously the plaquette occupation numbers are large for all values of $\mu$, and only a very mild change at $\mu_c \sim 1.35$ is visible. The flux variables show a small 
step at $\mu_c$ which, however, is less pronounced than in the strong coupling case.  Again we observe a strong splitting between the positive and negative temporal 
fluxes which reflects the influence of the chemical potential. At the largest values of $\mu$ we observe that the occupation numbers for positive and negative temporal 
flux approach each other in agreement with the physical picture developed on the discussion of Fig.~\ref{muplots}.

\section{Summary and discussion}

In this article we explored the possibility of a dual simulation of gauge theories with matter fields. Although our study is for a simple model, the Z$_3$ Gauge-Higgs model, 
it captures some of the features that
are expected also for more interesting theories, in particular the appearance of surfaces for the gauge fields and loops of flux for matter. In addition it was demonstrated that the complex action 
problem of the conventional representation at non-zero chemical potential is solved in the dual approach. 

A suitable Monte Carlo update was developed which properly treats the constraints of flux conservation at the sites and the surface constraints based at the links of the lattice (both constraints 
are modulo 3). In a detailed comparison in the pure gauge case and at vanishing chemical potential it was shown that the dual approach and the algorithm reproduce the 
results from a simulation in the conventional representation, thus establishing the validity of the dual approach. 

To test the approach at finite chemical potential $\mu$ where conventional techniques fail,  we explored the behavior of observables as a function of $\mu$ for two sets of couplings 
in the strong and weak coupling domains. In both cases we found first order transitions which were discussed not only based on usual observables, but also in terms of occupation numbers for 
the dual variables.  We stress at this point that our two case studies in the strong and weak coupling regimes do of course not constitute a systematic analysis of the phase diagram for the 
Z$_3$ Gauge-Higgs model -- a task we do not aim at here. 

The techniques developed in this paper can easily be generalized to other Gauge-Higgs systems with abelian groups \cite{alexandmeu1}.   The dual degrees of freedom are again surfaces 
and loops of flux, although the structure of the constraints and the weight factors will differ for other abelian groups. Some of the aspects and properties found in the Z$_3$ system might, 
however, turn out to be universal features of a dual approach:

\begin{itemize}

\item The dual formulation represents the system using only gauge invariant degrees of freedom, i.e., suitable occupation numbers for the plaquettes and the gauge invariant nearest neighbor 
terms of the matter fields.  

\item Part of the dynamics is encoded in expansion coefficients of the partition sum and the observables (e.g., the asymptotic $\mu$-behavior in the system studied here). 

\item Suitable Monte Carlo algorithms turn out to be rather simple, and at least for some cases a worm-type generalization to surfaces may be possible.

\item 
The dual representation is not only a tool to solve the complex action problem, but also allows for a conclusive discussion of the mechanisms at the various phase transitions
in terms of the dual variables. 

\end{itemize}

It is obvious that the current results only present first steps towards the more important cases of non-abelian gauge fields or systems with fermions. Nevertheless we expect that some of the techniques developed in the dual approach to abelian Gauge-Higgs systems might prove useful also for these more interesting cases.
 
\newpage
\noindent
{\bf Acknowledgements: } The authors would like to thank Shailesh Chandrasekharan, Ydalia Delgado Mercado,
Philippe de Forcrand, David Kaplan, Thomas Kloiber, Christian Lang and Anyi Li for interesting discussions and remarks on the literature. A.S.\ is funded by the FWF DK W1203 ''{\sl Hadrons 
in Vacuum, Nuclei and Stars}". C.G.\ acknowledges support from the Dr.~Heinrich J\"org foundation of the Karl-Franzens-University, Graz. C.G.\ also thanks the members of the
INT at the University of 
Washington, Seattle where part of this work was done for hospitality and support. This work is also partly supported by DFG TR55, {\sl Hadron Properties from Lattice QCD}.

\end{document}